\documentclass[useAMS,usenatbib,longnamesfirst,usedcolumn]{mn2e} 
\pdfoutput=1    
\usepackage{epsfig}	
\usepackage{times}	
\usepackage{pictex}
\usepackage{multirow}
\usepackage{subfigure}	
\usepackage[a4paper=true,pagebackref,
  pdftitle={The baryon budget on the galaxy group/cluster boundary},
  pdfauthor={Alastair J. R. Sanderson et al.},
  pdfsubject={MNRAS},
  pdfkeywords={keyword1 keyword2},
  pdfproducer={LaTeX with hyperref},
  pdfcreator={LaTeX},
  pdfdisplaydoctitle=true,
  colorlinks=true,
  citecolor=blue,
  linkcolor=Firebrick4,
  urlcolor=SteelBlue4]{hyperref}
\usepackage[hyperref,x11names]{xcolor}

\title{The baryon budget on the galaxy group/cluster boundary}
      \author[A. J. R. Sanderson et al.]{Alastair J. R. Sanderson$^{1}$\thanks{E-mail: ajrs@star.sr.bham.ac.uk},
        Ewan O'Sullivan$^{1, 2}$, Trevor J. Ponman$^1$,  \newauthor Anthony H. Gonzalez$^3$, Suresh Sivanandam$^{4, 5}$, Ann I. Zabludoff$^4$, Dennis Zaritsky$^4$\\
 $^1$School of Physics and Astronomy, University of
        Birmingham, Edgbaston, Birmingham B15 2TT, UK \\
 $^2$Harvard-Smithsonian Center for Astrophysics, 60 Garden Street,
        Cambridge, MA 02138\\
 $^3$Department of Astronomy, Bryant Space Science Center, University of Florida, Gainesville, FL 32611, USA \\
 $^4$Steward Observatory, University of Arizona, 933 North Cherry Avenue, Tucson, AZ 85721, USA\\
 $^5$Dunlap Fellow, Dunlap Institute for Astronomy and Astrophysics,
   University of Toronto, 50 St. George St, Toronto, ON M5S 3H4, Canada
       \\}
     \date{Accepted 2012 December 7. 
       Received 2012 December 5; 
       in original form 2012 October 22
       ($svn$ $Revision: 337 $)}

\pagerange{\pageref{firstpage}--\pageref{lastpage}}
\pubyear{2013}

\shortcites{balogh11,balogh08,biviano06,bower97b,connelly12,croston08,dai10,eckert12,ettori06,feldmeier02,hicks08,hou09,jansen01,kalberla05,kay04,maughan12,mccarthy11,mccarthy10,muanwong01,rasia06,rasmussen06b,rykoff08a,rykoff08b,san09b,san09,san03,sivanandam09,sun09,urban11,vanDaalen11,vikhlinin07,vikhlinin06,zibetti05}

\defcitealias{ascasibar08}{AD08}
\defcitealias{gonzalez07}{GZZ07}

\newcommand{\rmsub}[2]{\ensuremath{#1_{\mathrm{#2}}}} 
\newcommand{\srel}[2]{\mbox{\ensuremath{#1 - #2}}} 
\newcommand{\Chandra}{\emph{Chandra}}
\newcommand{\chisq}{\ensuremath{\chi^2}}

\newcommand{\fbary}{\rmsub{f}{b}}
\newcommand{\fstar}{\rmsub{f}{star}}
\newcommand{\fgas}{\rmsub{f}{gas}}

\newcommand{\km}{\ensuremath{\mbox{~km}}}
\newcommand{\kmpspMpc}{\ensuremath{\km \ps \pMpc\,}}

\newcommand{\LX}{\rmsub{L}{X}}
\newcommand{\Mfiveh}{\rmsub{M}{500}}
\newcommand{\Mgas}{\rmsub{M}{gas}}
\newcommand{\Mpc}{\ensuremath{\mbox{~Mpc}}}
\newcommand{\Msol}{\rmsub{M}{\odot}}
\newcommand{\MT}{\srel{M}{\TX}}

\newcommand{\pMpc}{\ensuremath{\Mpc^{-1}}}
\newcommand{\ps}{\ensuremath{\s^{-1}}}
\newcommand{\Rosat}{\textit{Rosat}}
\newcommand{\Rproject}{\textsc{r}}
\newcommand{\rhogasr}{\rmsub{\rho}{gas}\ensuremath{(r)}}
\newcommand{\rtwoh}{\rmsub{r}{200}}
\newcommand{\rfiveh}{\rmsub{r}{500}}
\newcommand{\rtwofiveh}{\rmsub{r}{2500}}

\newcommand{\s}{\ensuremath{\mbox{~s}}}
\newcommand{\Tbar}{\ensuremath{\overline{T}}}

\newcommand{\TX}{\rmsub{T}{X}}

\newcommand{\XMM}{\emph{XMM-Newton}}
\newcommand{\XSPEC}{\textsc{xspec}}

\begin{document}

\maketitle

\label{firstpage}

\begin{abstract}
  
 \noindent 
 We present a study of the hot gas and stellar content of 5
 optically-selected poor galaxy clusters, including a full accounting
 of the contribution from intracluster light (ICL) and a combined hot
 gas and hydrostatic X-ray mass analysis with \XMM\ observations. We
 find weighted mean stellar (including ICL), gas and total baryon mass
 fractions within \rfiveh\ of $0.026\pm0.003$, $0.070\pm0.005$ and
 $0.096\pm0.006$ , respectively, at a corresponding weighted mean
 \Mfiveh\ of ($1.08_{-0.18}^{+0.21}) \times
 10^{14}$\,M$_{\odot}$. Even when accounting for the intracluster
 stars, 4 out of 5 clusters show evidence for a substantial baryon
 deficit within \rfiveh, with baryon fractions (\fbary) between
 50$\pm$6 to 59$\pm$8 per cent of the Universal mean level
 (i.e. $\Omega_b / \Omega_m$); the remaining cluster having \fbary\ =
 75$\pm$11 per cent.  For the 3 clusters where we can trace the hot
 halo to \rfiveh\ we find no evidence for a steepening of the gas
 density profile in the outskirts with respect to a power law, as seen
 in more massive clusters. We find that in all cases, the X-ray mass
 measurements are larger than those originally published on the basis
 of the galaxy velocity dispersion ($\sigma$) and an assumed
 $\sigma-\Mfiveh$ relation, by a factor of 1.7--5.7. Despite these
 increased masses, the stellar fractions (in the range 0.016--0.034,
 within \rfiveh) remain consistent with the trend with mass published
 by \citet{gonzalez07}, from which our sample is drawn.
\end{abstract}



\section{Introduction}
\label{sec:intro}

The hot gas and stellar content of collapsed structures are key
elements in the baryon budget of the Universe.  Conducting a full
census of these baryons is particularly important given the likely
effects of non-gravitational heating and radiative cooling
\citep[e.g.][]{voit01} on the hot gas and star formation.  Both these
mechanisms raise the entropy of the intracluster medium (ICM), either
by heating it directly or else condensing out the lowest entropy gas
from the hot phase, but each produces a different balance of hot and
cold baryons. However, tracking all of the baryons is observationally
challenging, particularly in the lowest mass haloes, where feedback
may be more influential. In such cases, hot gas may have been
displaced to the group/cluster outskirts or even ejected from the
(progenitor) halo altogether \citep[e.g.][]{mccarthy11}. Moreover, a
substantial fraction of the cold baryons may be locked up in
intracluster stars, whose diffuse and low surface brightness emission
is very difficult to detect
\citep[e.g.][]{feldmeier02,zibetti05,gonzalez05,krick07}.

%
\begin{table*}
\centering
\begin{tabular}{l*{9}{c}}
\hline \\[-2ex]
Cluster name & RA$^{a}$ & Dec.$^{a}$ & Redshift & H\textsc{i} Column$^{b}$ & Obsid$^{c}$ & Observation Date & $t_{\mathrm{MOS1}}$ & $t_{\mathrm{MOS2}}$ & $t_{\mathrm{pn}}$ \\
     & (J2000)  & (J2000)    &       & ($\times10^{20}$ cm$^{-2}$) &             & & (ks) & (ks) & (ks)          \\
\hline\\[-2ex] 
 Abell 2401  & 21 58 22.5 & -20 06 15 & 0.0571 & 2.39 & 0555220101 & 2010 Oct 29 & 54.6 & 54.7 & 41.8 \\
 Abell 2955  & 01 56 59.8 & -17 02 23 & 0.0943 & 1.77 & 0555220201 & 2008 Aug 02 & 75.6 & 75.5 & 51.5 \\
 Abell S0296 & 02 46 51.5 & -42 22 41 & 0.0696 & 1.86 & 0555220301 & 2008 Dec 26 & 50.7 & 52.9 & 34.1 \\
\hline
\end{tabular}
\caption{Key properties of the 3 galaxy clusters with new
  \XMM\ data. $^{a}$Coordinates of the brightest cluster galaxy, on
  which the X-ray halo is centred. $^{b}$Galactic H\textsc{i} column,
  interpolated to the cluster centroid using the data of
  \citet{kalberla05}. $^{c}$\XMM\ observation identifier. The last 3
  columns are the effective exposure times for the MOS and pn
  detectors after flare filtering.  }
\label{tab:basicdata}
\end{table*}

The importance of the ICL in the baryon budget is the subject of
current debate, with different methods giving a range of estimates for
its contribution to the total cluster stellar light. For example,
direct measurements by \citet{krick07} suggest an ICL fraction of
6--22 per cent in the $r$ band, within 25 per cent of the virial
radius, while \citet{gonzalez05, gonzalez07} estimate that the
combined contribution of the brightest cluster galaxy and ICL amounts
on average to 33 per cent of the total stellar light within \rtwoh,
based on detailed $I$ band imaging of 24 clusters. In contrast,
\citet{mcgee10} find a $\sim$50 per cent contribution, based on
mapping of hostless type Ia supernovae. In one of the largest studies
of its kind, \citet{zibetti05} favour a lower contribution, of
$\sim$11 per cent within 500\,kpc, based on stacking Sloan Digital Sky
Survey (SDSS) images of 683 clusters. However, the use of stacking to
map the ICL will tend to bias the inferred measurements towards lower
values, if uncollapsed systems are included in the sample. The
relatively wide variation in ICL estimates may partly be due to
differences in sample selection. For example, the
\citeauthor{gonzalez05} study preferrentially includes systems with
a prominent brightest cluster galaxy (BCG).

On the other hand, the contribution of hot gas is better determined,
and represents the largest component of the baryon budget in
clusters. However, the best constraints are based on X-ray selected
(and often archival) samples
\citep[e.g.][]{vikhlinin06,arnaud07,sun09}, which are biased towards
haloes with more centrally concentrated hot gas
\citep[e.g.][]{rykoff08b,dai10,balogh11}. Furthermore, there is a
long-standing tendency for X-ray studies to target different clusters
to those best studied at optical/(near)-infrared wavelengths. A full
accounting of hot gas and stars (including ICL) in the \emph{same}
systems is therefore lacking, but is clearly essential to establish
the overall balance and uncover any missing baryons. This is best
achieved by securing X-ray observations of optically well-studied
clusters, given the difficulty of mapping the ICL directly.
 
The goal of this paper is, for the first time, to combine X-ray
observations of the hot gas in \emph{low-mass} clusters with a full
galaxy + ICL stellar mass analysis, using the former to measure both
the gas and total mass.  This approach calls for X-ray follow-up
observations of clusters for which the ICL analysis has already been
completed, given the difficulty of directly mapping diffuse optical
light. The low-mass scale is particularly interesting to study since
it encompasses the regime where cluster properties become increasingly
diverse and where there is a more equal balance between hot and cold
baryons \citep[e.g.][]{gonzalez07}.  Our clusters\footnote{We use the
  term `cluster' to refer to all the galaxy groups/poor clusters in
  our sample} were drawn from the optically-selected sample of
\citet[hereafter \citetalias{gonzalez07}]{gonzalez07}, with the aim of
targeting the lowest mass objects ($\sigma \la 500\km \ps$) with the
best quality optical data, for which the gas and total mass within
\rfiveh\ could be measured by \XMM\ in a reasonable exposure time
(\rfiveh\ being the radius enclosing an overdensity of 500 with
respect to the critical density).

\begin{figure*}
  \centering \subfigure{
    \includegraphics[width=5.5cm]{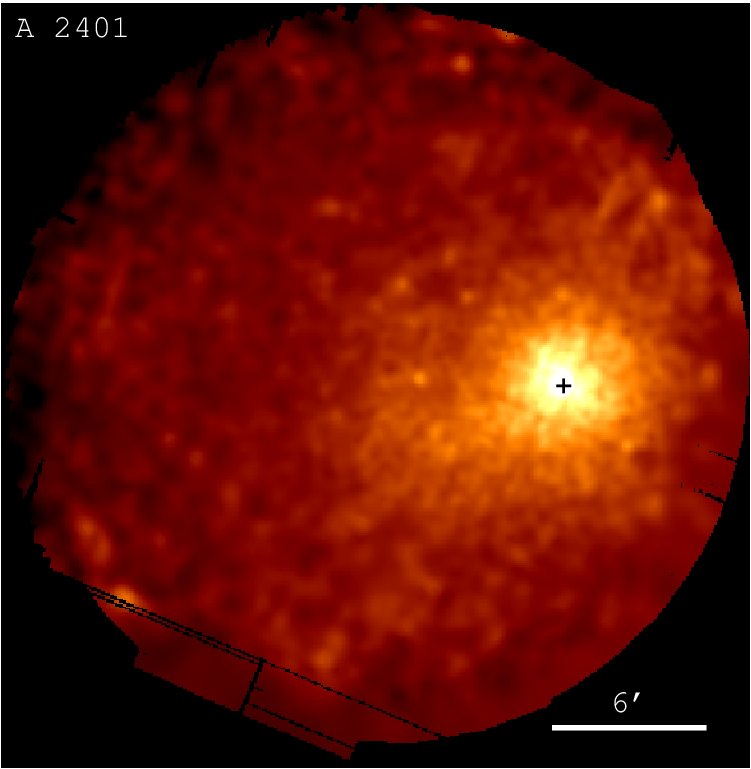}
  } \hspace{0cm} \subfigure{
    \includegraphics[width=5.5cm]{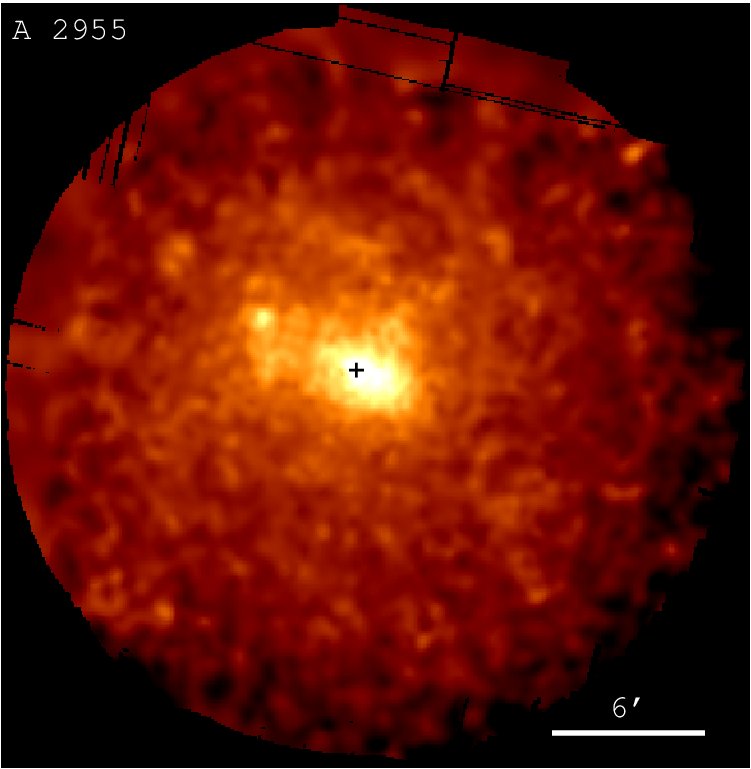}
  } \hspace{0cm} \subfigure{
    \includegraphics[width=5.5cm]{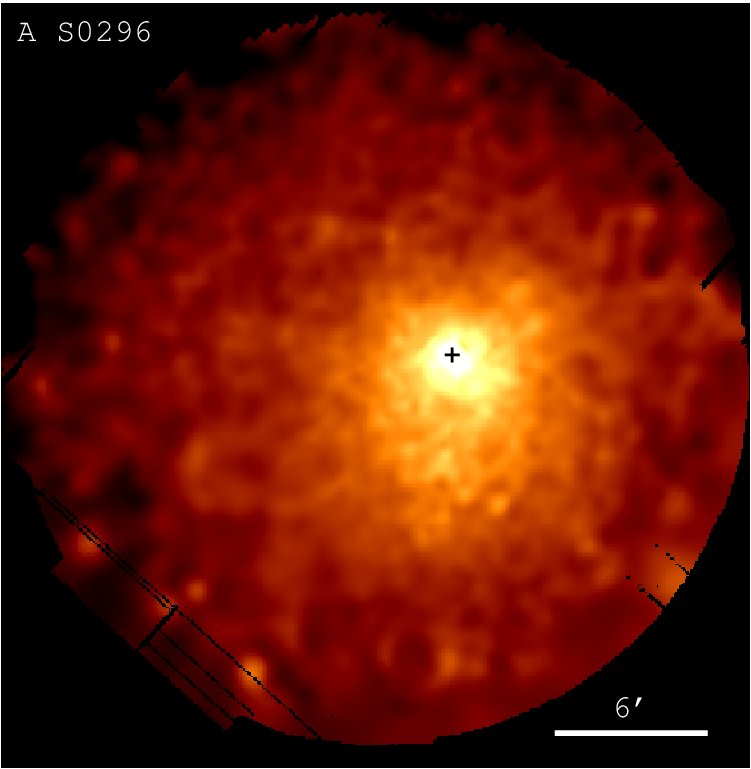}
  }
  \caption{Combined 0.3--2.0\,keV MOS/pn \XMM\ images of the diffuse
    emission from the 3 newly observed clusters, with point-like
    sources removed and filled using the \textsc{ciao} task
    \textsc{dmfilth}. The images have been adaptively smoothed (with
    $S/N=20$), particle background subtracted and exposure
    corrected. The position of the brightest cluster galaxy is marked
    with a `+'. Equivalent images of the 2 archival clusters can be
    found in \citet{sivanandam09}.}
  \label{fig:XMMimg}
\end{figure*}

Throughout this paper we adopt the following cosmological parameters:
$H_{0}=70$\kmpspMpc, $\rmsub{\Omega}{m}=0.3$ and
$\Omega_{\Lambda}=0.7$, converting any quoted data from other studies
to match this choice, where appropriate. Derived quantities scale with
the Hubble constant as follows: total mass, $M \propto h_{70}^{-1}$;
gas mass, $\rmsub{M}{gas} \propto h_{70}^{-5/2}$; gas fraction, $\fgas
\propto h_{70}^{-3/2}$; stellar mass, $\rmsub{M}{star} \propto
h_{70}^{-2}$; stellar fraction, $\fstar \propto h_{70}^{-1}$. Our
adopted value for the Universal baryon fraction is $\Omega_b h_{70}^2
/ \Omega_m h_{70}^2$ = $0.169 \pm 0.008$ from WMAP7
\citep{jarosik11}. For the conversion of $I$ band optical light to
stellar mass, we assume a mass-to-light ratio of 2.65 (see
Section~\ref{ssec:ICL_Mstar}). All errors are 1$\sigma$, unless
otherwise stated.

\section{X-ray data analysis}

\subsection{Data reduction and spectral profile deprojection}
\label{ssec:xmm_analysis}
The full sample of 5 clusters consists of three systems for which we
acquired new \XMM\ observations, plus a further two clusters drawn
from the archive (see Section~\ref{ssec:jfit_extra}). Our three
clusters were awarded time in \textit{XMM-Newton} Cycle~7, and
observed with the EPIC-pn operating in extended full frame mode and
the EPIC-MOS in full frame mode, with the thin optical blocking
filter. Observation dates and cleaned exposure times are given in
Table~\ref{tab:basicdata}. A detailed summary of the \XMM\ mission and
instrumentation can be found in \citet[and references
  therein]{jansen01}. Reduction and analysis were performed using the
\XMM\ Science Analysis System (\textsc{sas v.10.0.0}). We used the
\textsc{xmm-esas} analysis scheme, which is based on the background
modelling techniques described in \citet{snowden04}, which represents
the best currently-available method for analysing \XMM\ data.

The EPIC-MOS data were processed using the \textsc{emchain} and
\textsc{mos-filter} tasks, and the EPIC-pn data using \textsc{epchain}
and \textsc{pn-filter}. Bad pixels and columns were identified and
removed, and the events lists filtered to include only those events
with FLAG = 0 and patterns 0--12 (for the MOS cameras) or 0--4 (for
the pn). EPIC-MOS CCDs in anomalous states were excluded from further
analysis. These included Abell~2955 MOS1 CCD 4 and Abell~S0296 MOS1
CCDs 2 \& 4 and MOS2 CCDs 4 \& 5. Point sources were identified using
the \textsc{cheese-bands} task, including all three detectors, and
excluded from all further analysis. Sources associated with the centre
of the extended emission were considered to be mis-identified cuspy
cluster emission, rather than genuine point sources, and so were
ignored.

Since our goal was to determine the radial spectral profiles of the
clusters, we extracted spectra in annular regions centred on the peak
of the X-ray emission, which also coincides with the centroid of the
brightest cluster galaxy (BCG). The version of \textsc{esas} included
with \textsc{sas v.10.0.0} is only able to produce particle background
spectra for the EPIC-MOS cameras, so our spectral analysis was based
on these cameras only. However, we note that the background is lower
and better understood for the MOS than for the pn camera, which is
advantageous for the analysis of faint, extended emission. Moreover,
we were able to trace the X-ray haloes of our targets to larger radii
using \textsc{esas}, compared to a conventional MOS+pn analysis.

The \textsc{xmm-esas} scheme attempts to subtract the majority of the
particle component of the background using data taken with the
telescope filter wheels closed, and model the remaining X-ray and
particle components. Annuli were initially chosen to contain
2000--4000 net source counts, with a minimum bin width of 30\arcsec,
to ensure that the effects of blurring by the point spread function
are negligible. Where the source emission became too weak to meet this
criterion, the outer region of the field of view was divided into 2--3
large annuli, chosen to contain a sufficient number of counts to
provide well-defined background-only spectra. Spectra were extracted
using the \textsc{mos-spectra} and \textsc{mos\_back} tasks, and
grouped to contain at least 20 counts per bin.

Spectra were fitted in \XSPEC\ 12.6.0 following a procedure similar to
that described in the \textsc{xmm-esas}
cookbook\footnote{\url{ftp://xmm.esac.esa.int/pub/xmm-esas/xmm-esas.pdf}}.
Energies outside the range 0.3--10.0~keV were ignored and all EPIC-MOS
spectra were fitted simultaneously. An additional ROSAT All-Sky Survey
spectrum was extracted in an annulus between 0.5--1\degr\ from the
cluster centre, using the HEASARC X-ray Background
Tool\footnote{\url{http://heasarc.gsfc.nasa.gov/cgi-bin/Tools/xraybg/xraybg.pl}},
and was also included to help constrain the soft X-ray background.
The residual particle background in each MOS camera was modelled using
a power law whose index was tied across all annuli, with
normalizations tied between annuli in the ratio of the expected scaled
average flux, estimated using the \textsc{proton\_scale} task.
Similarly, the instrumental Al K$\alpha$ and Si K$\alpha$ fluorescence
lines were modelled using Gaussians whose widths and energies were
tied across all annuli, but with normalizations free.

The X-ray background was modelled as a combination of four components
whose normalizations were tied between annuli, scaling to a
normalization per square arcminute as determined by the
\textsc{proton-scale} task. The cosmic hard X-ray background was
modelled using a power law with index $\Gamma=1.46$ and normalization
$=8.88\times10^{-7}$ photons keV$^{-1}$cm$^{-2}$s$^{-1}$ at 1~keV. The
local soft X-ray background was modelled by a 0.1~keV unabsorbed
\textsc{apec} model representing the local hot bubble, an absorbed
0.1~keV \textsc{apec} model representing the cool galactic halo, and a
0.25~keV \textsc{apec} model representing the hotter component of the
galactic halo or the intergalactic medium. The relative normalizations
of the \textsc{apec} models were free to vary, and the temperature of
the hottest component was also allowed to vary after the initial
fit. The line of sight absorption was modelled using the \textsc{wabs}
component, fixed to the galactic column density taken from the
\textsc{ftools} task \texttt{nh} \citep[based on the data
  of][]{kalberla05}. The RASS spectrum was fitted using only these
X-ray background components.

The contribution of the cluster was modelled using a deprojected
(using the \textsc{projct} component) absorbed \textsc{apec}
model. Temperature, abundance and normalization were free to vary in
each bin, and where the best fitting normalization approached zero (in
the outer bins) it was fixed to zero. Abundances were measured
relative to the abundance ratios of \citet{grevesse98}. Where values
of abundance (or in some cases temperature) were poorly constrained,
or where `ringing' between bins was observed, parameter values were
tied between pairs of bins. The resulting best-fitting values of
normalization and 1$\sigma$ errors were then used to calculate the gas
density. Finally, we performed a cross-check of our background
modelling in the outermost bin of A2955 (which is free of emission
from the cluster halo), and found that the 0.3--10\,keV count rate
predicted by the background component of the model was within 1.5 and
0.7 per cent of the measured rate for MOS1 and MOS2, respectively.

 \subsection{Cluster modelling}
 \label{ssec:clusmodel}
 In order to determine the total mass and gas fraction of the
 clusters, we fitted the deprojected gas temperature and density
 profiles with the phenomenological cluster model of
 \citet{ascasibar08}, following the method described in
 \citet{san10}. This is a physically-motivated, parsimonious yet
 flexible model, in which the ICM is modelled as a polytropic gas in
 hydrostatic equilibrium within a \citet{hernquist90} gravitational
 potential, with a variable cool-core component. It has been shown to
 provide a good description of the temperature and density profiles of
 a wide variety of galaxy clusters, fully encompassing the mass range
 of the objects studied here. \citep{san09b,san10}. We use mean
 molecular weight of 0.601 atomic mass units, based on a fully ionized
 plasma with a mean metallicity of 0.5 Solar and using the
 \citet{grevesse98} \textsc{xspec} abundance
 table\footnote{\url{http://heasarc.gsfc.nasa.gov/docs/xanadu/xspec/manual/XSabund.html}}. The
 associated ratio of electron to hydrogen (strictly, the atomic mass
 unit) number density is 1.157, used in calculating the gas mass. A
 characteristic radius was assigned to each annulus using the
 emission-weighted approximation of \citet{mcl99}:
 \begin{equation}
 r = \left[ 0.5\left( \rmsub{r}{out}^{3/2} + \rmsub{r}{in}^{3/2} \right)
 \right]^ {2/3} .
 \end{equation}

\begin{figure*}
   \includegraphics[width=17.5cm]{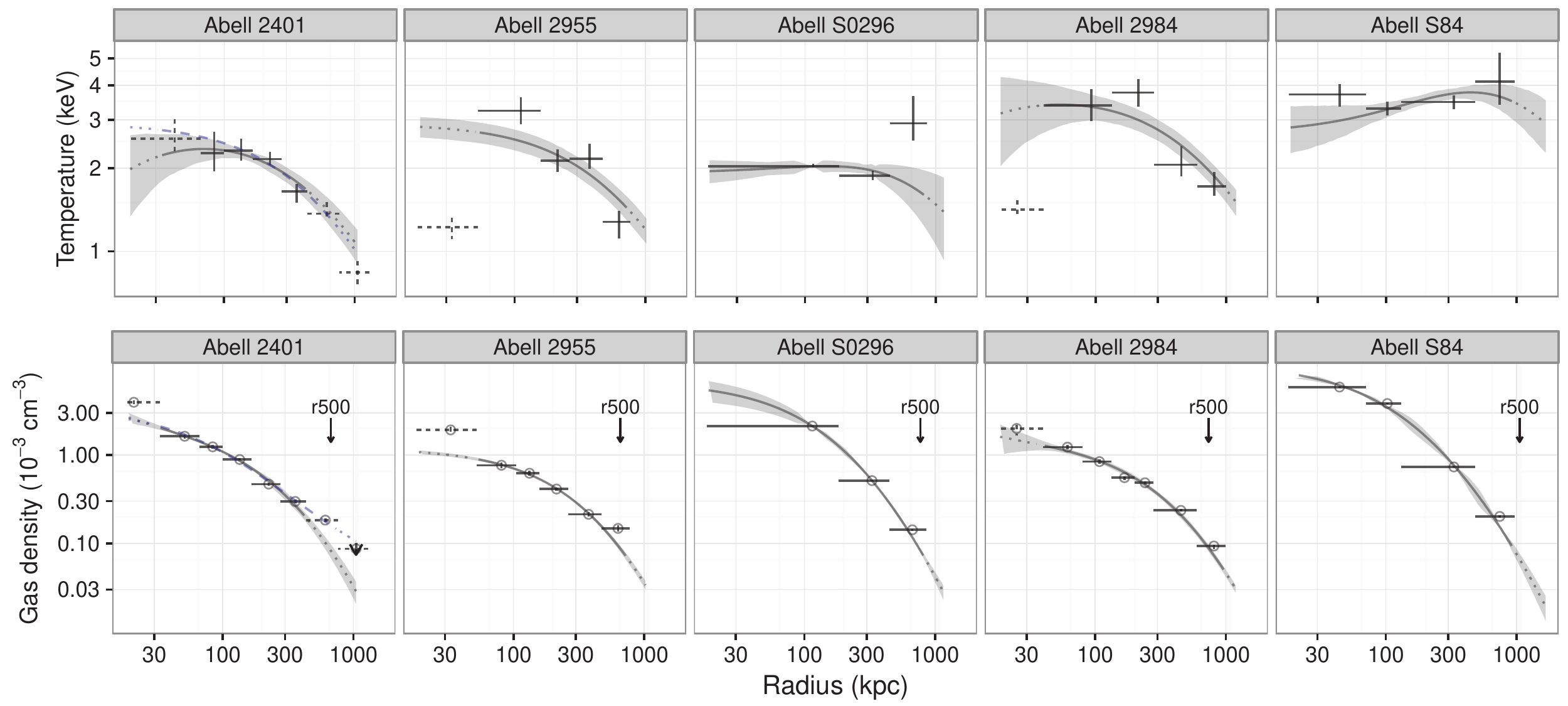}
   \caption{ \label{fig:jfit_kT_rhogas} Gas temperature and density
     profiles for the clusters (error bars \& hollow circles + error
     bars, respectively) with the best-fit \citetalias{ascasibar08}
     cluster model (solid line) and 1$\sigma$ error envelope. Also
     shown, in columns 4 and 5, are two additional clusters included
     to increase the sample size, as described in
     section~\ref{ssec:jfit_extra}. The innermost bin was excluded
     from the fit in all cases and is plotted with dashed error
     bars. Dotted curves indicate extrapolation of the model beyond
     the data and the blue dashed curve for A2401 is the (much worse)
     best-fit model including the penultimate bin and with the
     polytropic index free to vary. Note that for A2401 the outermost
     bin is excluded in both cases and so the corresponding density
     point has not been corrected for projection and is plotted as an
     upper limit (see text for details).}
 \end{figure*}

 The cluster model was jointly fitted to the deprojected gas density
 and temperature profiles, excluding the central bin, since our focus
 is on global properties and we expect complex baryon physics on the
 scale of the BCG to invalidate the model assumptions here; the fitted
 data and model are plotted in Fig.~\ref{fig:jfit_kT_rhogas}. To
 improve the radial coverage of the gas distribution, we also use the
 cluster model to determine a correction factor for estimating the
 density in the outermost bin. This point is often omitted on the
 basis that the emission in the outer annulus has not been deprojected
 and so includes a contribution from gas distributed along the line of
 sight beyond the radius of the corresponding spherical
 shell. However, we can use the cluster model to estimate what
 fraction of the total emission in this annulus originates from the
 corresponding spherical shell.

\begin{table*}
\begin{center}
\begin{tabular}{cccccccccccc}
  \hline
Cluster & $T_0$\,(keV) & $t$ & $a$ (kpc) & $\alpha$ & $f$ & \rfiveh\ (kpc) & $r_{\mathrm{fit}}$ (kpc) & \chisq\ / dof & $p$ value & $f_{\rho}$ & $f_{kT}$ \\ 
  \hline
Abell 2401 & 2.94\,$\pm$\,0.25 & 0\,$\pm$\,0.11 & 588\,$\pm$\,135 & 0.01\,$\pm$\,0.03 & 0.73\,$\pm$\,0.18 & 672\,$\pm$\,20 & 432 & 13.3/4 & 0.010 & 0.03 & 0.02 \\ 
  Abell 2955 & 2.9\,$\pm$\,0.30 & 1\,$\pm$\,0.00 & 709\,$\pm$\,51 & 0.01\,$\pm$\,0.01 & 0.78\,$\pm$\,0.11 & 656\,$\pm$\,32 & 770 & 13.0/4 & 0.011 & 0.01 & 0.13 \\ 
  Abell S0296 & 3.91\,$\pm$\,0.38 & 0\,$\pm$\,0.00 & 750\,$\pm$\,125 & 0.01\,$\pm$\,0.02 & 0.73\,$\pm$\,0.12 & 772\,$\pm$\,39 & 997 & 36.8/5 & 0.000 & 0.07 & 0.06 \\ 
   \hline
Abell 2984 & 3.72\,$\pm$\,1.06 & 0.52\,$\pm$\,0.16 & 870\,$\pm$\,382 & 0.49\,$\pm$\,0.48 & 0.30\,$\pm$\,0.18 & 738\,$\pm$\,66 & 861 & 10.6/1 & 0.001 & 0.00 & 0.23 \\ 
  Abell S84 & 7.69\,$\pm$\,1.43 & 0.35\,$\pm$\,0.07 & 1337\,$\pm$\,337 & 0.38\,$\pm$\,0.69 & 0.18\,$\pm$\,0.08 & 1054\,$\pm$\,85 & 963 & 52.4/3 & 0.000 & 0.06 & 0.06 \\ 
   \hline
\end{tabular}
\caption{Best-fitting model parameters for each cluster and 1$\sigma$ errors estimated from 200 bootstrap resamples. $r_{\mathrm{fit}}$ is the radius of the outer annulus of the fitted data in the spectral profile. The last four columns are the \chisq\ / degrees of freedom with associated $p$ value and the mean fractional intrinsic scatter in the gas density and temperature, respectively \citep[as defined in][]{san10}. The lower portion of the table refers to two additional clusters included to increase the sample size, as described in section~\ref{ssec:jfit_extra}.}
\label{tab:jfitpars}
\end{center}
\end{table*}

To do this, we consider a line of sight column located at the
half-radius of the spherical shell, and intersecting its outer edge at
two points. The correction factor is simply the ratio of the density
squared integrated along this line within the shell to that evaluated
between $\pm$ infinity\,-- this weighting approximates the emissivity of
bremsstrahlung as well as that of low-temperature line emission. This
fraction is then multiplied by the density obtained by assuming
\emph{all} the flux in the outer annulus is associated with the volume
of the corresponding spherical shell. With the outer density point
thus determined, a new best-fit model is obtained and used to evaluate
a corresponding new estimate of the correction factor, until
convergence is achieved. The final correction factors obtained are
0.70 for A2955 and 0.73 for AS0296. This then allows the outermost density
point to be used in the analysis, albeit at the expense of introducing
some model-dependency.

\begin{table*}
\begin{center}
\begin{tabular}{cccccccc}
  \hline
Cluster & $M_{2500}$\,($10^{14} M_{\sun}$) & $M_{500}$\,($10^{14} M_{\sun}$) & \rtwofiveh\,(kpc) & \rfiveh\,(kpc) & \fgas ($<$\rtwofiveh) & \fgas ($<$\rfiveh) & $\alpha^{a}$ \\ 
  \hline
Abell 2401 & 0.34\,$\pm$\,0.05 & 0.91\,$\pm$\,0.08 & 282\,$\pm$\,15 & 672\,$\pm$\,20 & 0.047\,$\pm$\,0.003 & 0.073\,$\pm$\,0.009 & -0.48\,$\pm$\,0.17 \\ 
  Abell 2955 & 0.28\,$\pm$\,0.06 & 0.88\,$\pm$\,0.13 & 260\,$\pm$\,17 & 656\,$\pm$\,32 & 0.036\,$\pm$\,0.002 & 0.064\,$\pm$\,0.005 & -0.14\,$\pm$\,0.01 \\ 
  Abell S0296 & 0.48\,$\pm$\,0.12 & 1.4\,$\pm$\,0.21 & 315\,$\pm$\,27 & 772\,$\pm$\,39 & 0.039\,$\pm$\,0.003 & 0.065\,$\pm$\,0.008 & -0.29\,$\pm$\,0.25 \\ 
   \hline
Abell 2984 & 0.37\,$\pm$\,0.07 & 1.3\,$\pm$\,0.35 & 286\,$\pm$\,20 & 738\,$\pm$\,66 & 0.087\,$\pm$\,0.004 & 0.096\,$\pm$\,0.014 & -0.32\,$\pm$\,0.10 \\ 
  Abell S84 & 1\,$\pm$\,0.18 & 3.7\,$\pm$\,0.96 & 399\,$\pm$\,23 & 1054\,$\pm$\,85 & 0.080\,$\pm$\,0.015 & 0.075\,$\pm$\,0.019 & -0.47\,$\pm$\,0.11 \\ 
   \hline
\end{tabular}
\caption{Derived quantities from the cluster models with 1$\sigma$ errors estimated from 200 bootstrap resamples. $^a$Logarithmic gradient of the gas density profile at 0.04\rfiveh\ \citep[after][]{vikhlinin07}; $\alpha\la -0.85$ indicates a strong cool core \citep{san09b}. The lower portion of the table refers to two additional clusters included to increase the sample size, as described in section~\ref{ssec:jfit_extra}.}
\label{tab:derivdata}
\end{center}
\end{table*}

For A2401, the two outermost annular bins are only partially covered
by the \XMM\ CCD chips, as a result of the off-axis position of the
cluster halo (see left panel of Fig.~\ref{fig:XMMimg}), resulting in
systematically incomplete azimuthal coverage. For the outermost bin
only 23 per cent of the annulus falls on the chip, while the
penultimate bin has a coverage of 59 per cent. Recent work by
\citet{eckert12} has found substantial azimuthal variation in ICM
density in cluster outskirts, as a result of structural asymmetries
and gas clumping, which could introduce significant systematic bias
into the determination of the gas properties from incomplete outer
annuli. Moreover, both of these bins include emission from beyond
\rfiveh\ and even beyond \rtwoh\ for the outermost annulus, where the
intracluster gas increasingly deviates from hydrostatic equilibrium
\citep[e.g][]{nagai07} and may also be subject to complex physics like
clumping \citep[e.g. see][and references therein]{urban11}, which
would invalidate the model. We therefore exclude both the outer two
bins from the fit and so are unable to determine a corrected density
value for the outermost annulus, as this is model-dependent. However,
since the uncorrected density represents an upper limit, we plot it in
Fig.~\ref{fig:jfit_kT_rhogas} with an arrow extending from the
1$\sigma$ upper bound.

The best-fit parameters for the final cluster models and their
1$\sigma$ errors (estimated from 500 bootstrap re-samples) are listed
in Table~\ref{tab:jfitpars}, together with values of \rfiveh\ and the
radius of the outermost annular bin included in the fit. Following
\citet{san10}, we consider the residual fractional intrinsic scatter
of the gas density data about the model (which measures the real
variance of the data about the model, beyond that expected from the
measurement errors) as a useful indicator of the effectiveness of the
model in describing the data (second last column of
Table~\ref{tab:jfitpars}). This quantity is the mean value of $\left[
  (y_{obs} - y_{pred})^2 - y_{err}^2\right] / y_{pred}$, with all
negative values to zero, based on the observed values, predicted
values and measurement errors of $y$ \citep[see the appendix
of][]{san10}. For A2401 and A2955 this scatter is three and one per
cent, respectively, in gas density-- well below the levels of $>$5 per
cent associated with merging/disrupted clusters for which the model
performs less well \citep{san10}. However, AS0296 has a 7 per cent
residual intrinsic scatter in density, which hints at a somewhat
irregular gas distribution which the model does not capture well. The
fractional intrinsic scatter for the gas temperature is also presented
in Table~\ref{tab:jfitpars}, although this is a less reliable
diagnostic of model effectiveness \citep{san10}.

According to the best-fit models, none of these clusters hosts a
sharply peaked cool core: the inner cuspiness as measured by the
logarithmic slope of the gas density at 0.04\rfiveh\ \citep[$\alpha$,
cf.][]{vikhlinin07,san09b} is flatter in all cases than the value of
$\la -0.85$ which typically characterizes strong cool core clusters
\citep[see Table~\ref{tab:derivdata}]{san09b}. On the other hand, it
is noticeable from Fig.~\ref{fig:jfit_kT_rhogas} that the (excluded)
innermost temperature point (radius $\la$\,40--50\,kpc) lies
substantially below the extrapolated model prediction in the cases of
A2955 and AS0296. Our Ascasibar \& Diego cluster model is fully able
to describe cool cores \citep[cf.][]{ascasibar08,san10}, but
hydrostatic equilibrium is built into this model, and so deviations
are expected on the scale of the central galaxy. In any case, the
possible presence of small cool cores in these two clusters has no
significant impact on the present study, which is concerned with the
large scale properties of these systems.

\subsection{Analysis of two extra poor clusters}
\label{ssec:jfit_extra}
In order to increase our coverage of the baryon fraction in
groups/poor clusters for which a full intracluster light analysis has
been performed, we have added two extra systems to our sample from a
companion paper (Abell 2984 \& Abell~S84; Gonzalez et al., in
preparation). These objects were also part of the
\citetalias{gonzalez07} sample and their \XMM\ data were analysed in
\citet{sivanandam09}. For consistency, we have re-analysed these poor
clusters in the same way as for our other three systems (i.e. as
described in Sections~\ref{ssec:xmm_analysis} \&
\ref{ssec:clusmodel}), and their corresponding model parameters and
derived quantities are listed in the lower portions of
Tables~\ref{tab:jfitpars}~\&~\ref{tab:derivdata} . The corresponding
correction factors for determining the gas density in their outermost
bins are 0.83 \& 0.82 for A2984 \& AS84, respectively. We address the
comparison between our modelling analysis of these systems and that of
\citet{sivanandam09} in Sections~\ref{ssec:M500_check} \&
\ref{ssec:betamodel}.

\section{Optical data}
\label{sec:optical}
Cluster membership was determined on the basis of galaxy velocities,
which were obtained from dedicated spectroscopic (and associated
photometric) observations in the case of A2955 and AS0296, yielding a
total of 22 and 34 cluster members, respectively \citep{zaritsky06}
. These galaxies were originally selected for spectroscopic follow-up
if they were both fainter than the BCG and located within a projected
distance of 1.5\,Mpc of it, and the resulting velocity distribution
was subjected to a 3$\sigma$ clipping algorithm, to select the cluster
members \citep{zaritsky06}, using biweight estimators of location and
scale for improved robustness \citep{beers90}.

\begin{figure*}
\includegraphics[width=15cm]{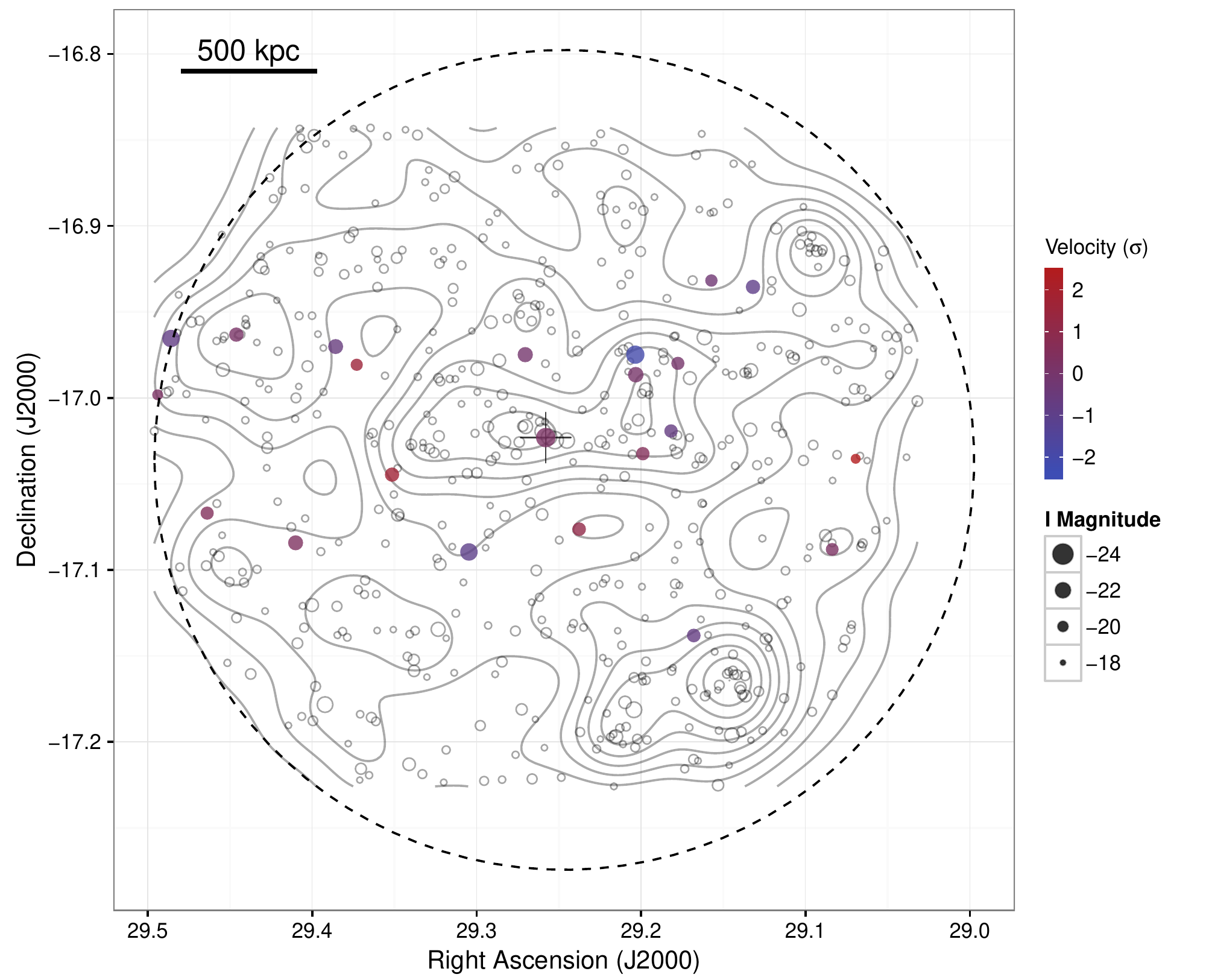}
\caption{ \label{fig:radec-a2955} Positions of galaxies around
  Abell~2955 with corresponding isodensity contours superposed,
  Gaussian smoothed on a scale of 500\,kpc. The point sizes indicate
  the $I$ band absolute magnitude (assuming all the galaxies are at
  the redshift of the cluster) and the filled points are the cluster
  members, coloured according to the velocity with respect to the
  cluster mean velocity, normalized by the velocity dispersion (see
  Table~\ref{tab:optdata}). The position of the brightest cluster
  galaxy is marked with a `+' and the dashed circle marks the
  field-of-view of the corresponding \XMM\ observation.}
\end{figure*}

\begin{table*}
\begin{center}
\begin{tabular}{ccccccccc}
  \hline
Cluster & \fstar ($<$\rtwofiveh) & \fstar ($<$\rfiveh) & \fbary ($<$\rtwofiveh) & \fbary ($<$\rfiveh) & $M_{500}$ ($\sigma$)\,($10^{14} M_{\sun}$) & $\sigma$ (km/s) & N$_{\mathrm{gal}}$ & BM Type \\ 
  \hline
Abell 2401 & 0.036\,$\pm$\,0.006 & 0.028\,$\pm$\,0.003 & 0.083\,$\pm$\,0.006 & 0.101\,$\pm$\,0.013 & 0.55 & 459$_{-83}^{+101}$ & 24 & II \\ 
  Abell 2955 & 0.064\,$\pm$\,0.013 & 0.034\,$\pm$\,0.005 & 0.100\,$\pm$\,0.006 & 0.098\,$\pm$\,0.009 & 0.15 & 316$_{-67}^{+85}$ & 22 & II \\ 
  Abell S0296 & 0.035\,$\pm$\,0.009 & 0.019\,$\pm$\,0.003 & 0.074\,$\pm$\,0.007 & 0.085\,$\pm$\,0.010 & 0.63 & 477$_{-65}^{+76}$ & 34 & I \\ 
   \hline
Abell 2984 & 0.073\,$\pm$\,0.014 & 0.031\,$\pm$\,0.009 & 0.159\,$\pm$\,0.008 & 0.127\,$\pm$\,0.019 & 0.69 & 490$_{-91}^{+112}$ & 29 & I \\ 
  Abell S84 & 0.035\,$\pm$\,0.006 & 0.016\,$\pm$\,0.004 & 0.115\,$\pm$\,0.022 & 0.091\,$\pm$\,0.023 & 0.85 & 522$_{-122}^{+155}$ & 24 & I \\ 
   \hline
\end{tabular}
\caption{Optically-derived data for the clusters. The last 4 columns are the optically-derived mass, velocity dispersion, number of galaxies with velocities and Bautz-Morgan type, repectively \citep[from][\citetalias{gonzalez07}]{zaritsky06}. The lower portion of the table refers to two additional clusters included to increase the sample size, as described in section~\ref{ssec:jfit_extra}.}
\label{tab:optdata}
\end{center}
\end{table*}

\begin{figure*}
\includegraphics[width=15cm]{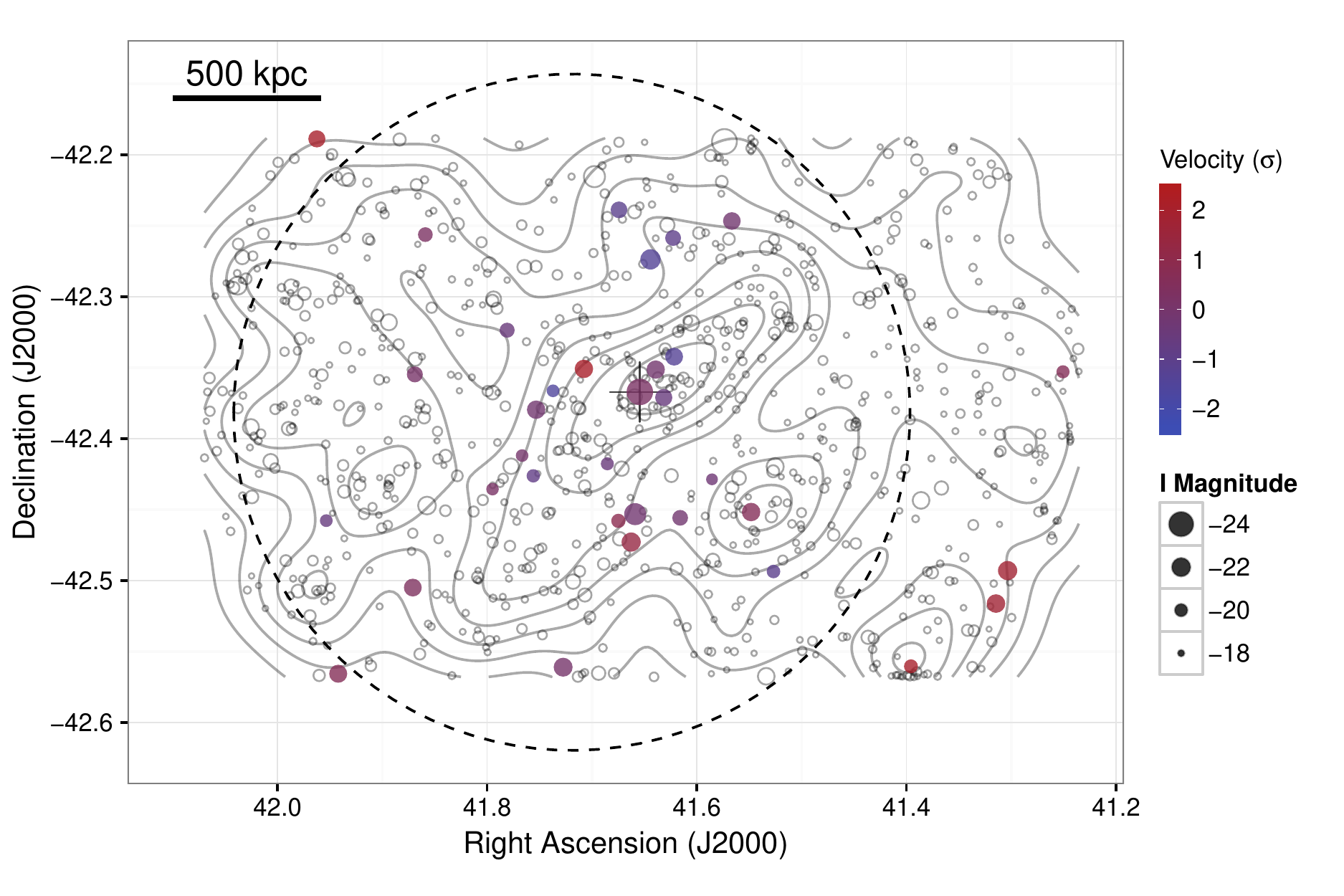}
\caption{ \label{fig:radec-aS0296} As for Fig.~\ref{fig:radec-a2955}, but
 for Abell S0296.}
\end{figure*}

\begin{figure*}
\includegraphics[width=12cm]{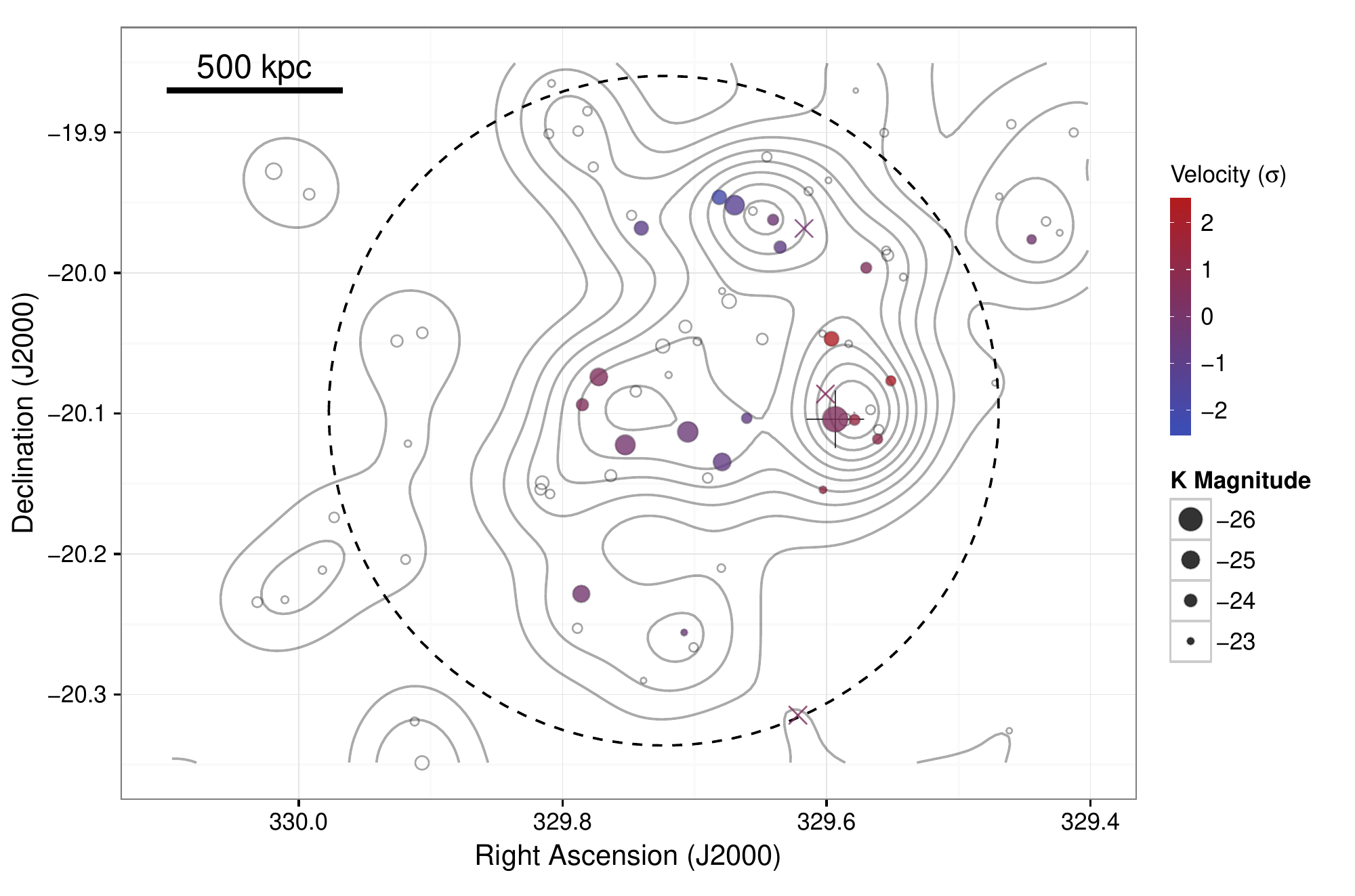}
\caption{ \label{fig:radec-a2401} As for Fig.~\ref{fig:radec-a2955},
  but for Abell~2401 and using 2MASS $K$ band magnitudes. In this case
  the cluster members (filled, coloured points) were selected using
  data from the NED and the `$\times$' symbols mark the positions of 3
  galaxies with velocities but no corresponding $K$ magnitude.}
\end{figure*}

The positions of galaxies are plotted in Figs.~\ref{fig:radec-a2955}
\& \ref{fig:radec-aS0296} with point size indicating their $I$ band
absolute magnitude and with cluster members colour-coded according to
their relative line-of-sight velocity (hollow points have unknown
velocity). In addition, confirmed non-cluster members have been
omitted-- a total of 35 and 11 galaxies for A2955 and AS0296,
respectively. Overlaid are contours of surface number density which
were computed by kernel smoothing the positions of all the galaxies in
the field with a Gaussian of bandwidth 500\,kpc at the redshift of the
cluster, using the task \textsc{kde2d} in the \textsc{mass} package
\citep{venables02} in
\Rproject\footnote{\url{http://www.r-project.org}}. For A2401, galaxy
velocities were extracted from the NASA Extragalactic Database (NED),
giving a total of 24 cluster members. These were cross-matched with
the 2 Micron All Sky Survey (2MASS) Extended Source Catalogue
\citep[XSC;][]{skrutskie06} to provide $K$ band magnitudes and their
positions are similarly plotted in colour in
Fig.~\ref{fig:radec-a2401}, together with those of other 2MASS XSC
objects in the vicinity, for which the velocities are unknown
(coloured in grey).

In all cases the BCG lies very close to the peak of the number density
contours smoothed on a scale of 500\,kpc. However, other localized
galaxy concentrations are evident, particularly in the case of A2401,
which suggest either the presence of substructures in the cluster or
ongoing merging/disruption. Moreover, the density contours are based
on all the galaxies detected along the line of sight, not just those
with redshifts and that were judged to be cluster members. Therefore,
some of the features seen could be the result of unrelated fore- or
background structures.

\begin{figure*}
\includegraphics[width=17cm]{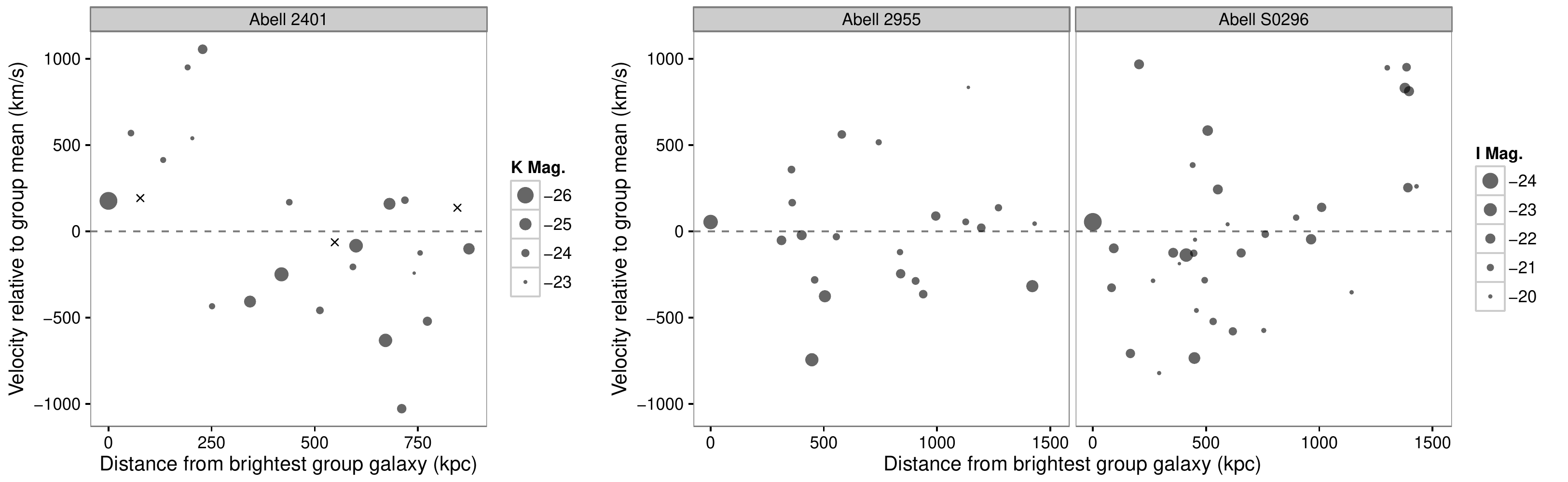}
\caption{ \label{fig:vel-radius} Line-of-sight velocity vs. distance
  from the brightest cluster galaxy for the galaxies spectroscopically
  confirmed as cluster members, point size coded by absolute
  magnitude. For Abell~2401, `$\times$' symbols mark the positions of
  3 galaxies with velocities but no corresponding $K$ magnitude.}
\end{figure*}

Fig.~\ref{fig:vel-radius} shows the variation in line-of-sight
velocity with distance from the BCG for galaxies spectroscopically
confirmed to be members of the cluster. The behaviour seen is broadly
consistent with the expected distribution for a virialized population
of tracer particles, with no major anomalies that might cast doubt on
the use of $\sigma$ as a mass estimator. For AS0296, there are four
galaxies in the top right corner of the plot which are likely to be
interlopers lying outside the virialized volume of the cluster; indeed
3 of these are visible as a clump of red dots in the lower right
corner of the galaxy distribution plot in Fig.~\ref{fig:radec-aS0296},
suggesting an infalling substructure. However, their impact on the
overall velocity dispersion of the cluster is minor and excluding
their contribution would only further exacerbate the mass discrepancy
compared to the larger X-ray determined value, discussed in
Section~\ref{ssec:Mx_vs_Mopt}. A2401 may also be affected by one or
two interlopers, on the basis of Fig.~\ref{fig:vel-radius}, but here
also any bias is likely to be small.

We performed several tests of normality of the galaxy velocity
distribution, using the \Rproject\ package \textsc{nortest}
\citep[e.g.][]{thode02}. For all three clusters, none of the tests
appropriate for these sample sizes shows evidence for a non-Gaussian
distribution, including the Anderson-Darling test, which has been
shown to be the most sensitive to optical cluster substructure
\citep{hou09}.

A key element in determining the stellar mass in clusters is the
aperture within which the measurement is made, which depends on the
total gravitating mass. In the original \citetalias{gonzalez07}
analysis, X-ray data were generally not available and so the total
mass (within \rfiveh) was estimated from the velocity dispersion of
the galaxies, using the relation between $\sigma$ and \Mfiveh\ based
on the cluster sample of \citet{vikhlinin06}, where a full hydrostatic
mass analysis was performed. There is evidence that $\sigma$ can often
underestimate the total mass in galaxy groups
\citep{osmond04,connelly12}, and so a key aim of the present study is
to determine the mass independently, using our X-ray data. Indeed, as
shown by \citet{balogh08}, the four lowest mass
\citetalias{gonzalez07} clusters, including A2955 from our study, have
uncomfortably high baryon fractions (implying an underestimated total
mass) on the basis of their velocity dispersions-derived
masses. However, these systems had been excluded by
\citetalias{gonzalez07} on the grounds that their mass estimates
required an extrapolation of the $\Mfiveh-\sigma$ relation and were
hence unreliable (we emphasize that they were \emph{not} excluded on
the basis of \fbary). We address the differences between the X-ray and
optically derived mass estimates in Section~\ref{ssec:Mx_vs_Mopt}.

\subsection{Intracluster light and stellar mass}
\label{ssec:ICL_Mstar}
\citetalias{gonzalez07} made drift-scan observations to measure the
optical light profile of the clusters, with accurately measured
background and flat-fields, in order to map the diffuse ICL directly
and with good precision in each case. The cluster radial light profile
was modelled as a combination of two de Vaucouleurs profiles, to
capture the separate contributions from the BCG and ICL. The total
optical luminosity of the cluster is calculated by integrating the
model within a given aperture and then adding in the luminosities of
the individual non-BCG cluster galaxies in the same aperture
(corrected for contamination).  Since our X-ray derived masses
(Table~\ref{tab:derivdata}) are substantially larger than the masses
derived from velocity dispersions by \citetalias{gonzalez07}, as we
will discuss in Section~\ref{ssec:Mx_vs_Mopt}, the apertures
corresponding to \rtwofiveh\ and \rfiveh, within which the optical
luminosity are integrated are correspondingly larger.  We therefore
re-measured the cluster optical luminosity (including ICL) within
these new radii. 

These aperture measurements represent the \emph{projected} luminosity
within a cylinder along the line of sight through the cluster. We have
therefore corrected them for projection, to calculate the 3
dimensional luminosity within a sphere, assuming a \citet{navarro95}
profile with a concentration parameter of 2.9 \citep[based on the
stacked profile of][]{lin04}. This correction only applies to the
galaxy distribution and not the BCG+ICL light, which is
(approximately) completely contained within the smallest aperture
considered here (\rtwofiveh). The ratios of the deprojected to
projected luminosities are in the range 0.74--0.91 within \rtwofiveh\
and 0.78--0.86 within \rfiveh.

Our adopted average $I$ band mass to light ratio (M/L) is 2.65, which
is lower than that used in \citetalias{gonzalez07}. This new value
incorporates a 15 per cent contribution from dark matter as well as a
correction to the previous \citetalias{gonzalez07} M/L calculation
which is described in more detail in a forthcoming paper (Gonzalez et
al, in preparation). The corresponding stellar masses and mass
fractions are summarized in Table~\ref{tab:optdata}, together with
other optical data for the 5 clusters. In
Section~\ref{ssec:other_syserrs} we address the impact on our results
of systematic errors in M/L.

\section{Results}
\label{sec:results}

\subsection{Radial trends in stellar and gas fraction}
\label{ssec:radial_trends}
\begin{figure*}
\includegraphics[width=17cm]{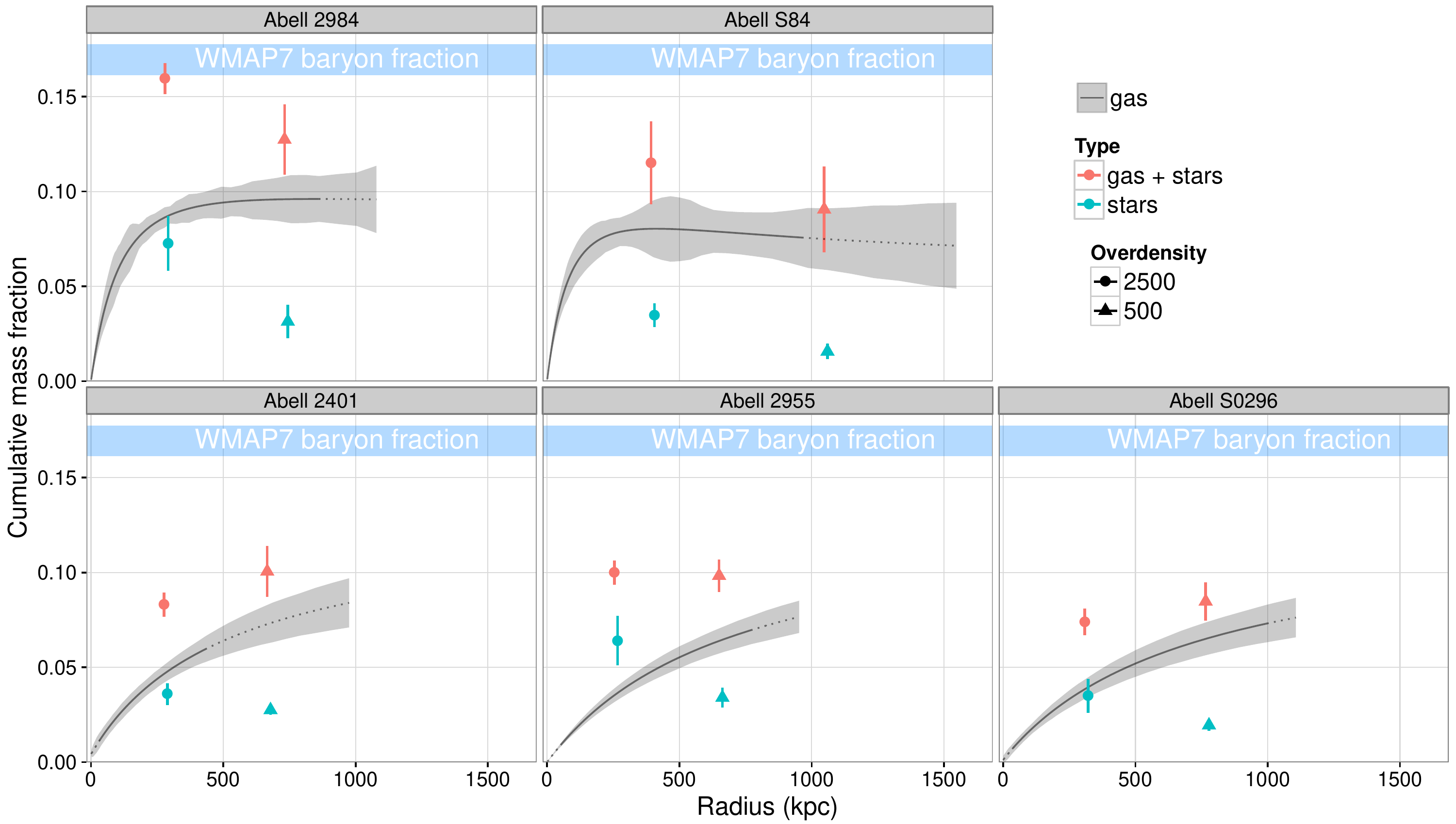}
\caption{ \label{fig:fbary(r)} Gas mass fraction profiles plotted out
  to \rtwoh\ for each cluster, with the stellar and total baryon
  fraction (including ICL) evaluated at \rfiveh\ and
  \rtwofiveh\ (offset slightly in radius to avoid over plotting). The
  grey region is the 1$\sigma$ confidence envelope on the gas fraction
  profile from the cluster model and a dotted curve denotes
  extrapolation of the model beyond the radius of the outermost bin
  (see Table~\ref{tab:jfitpars}).}
\end{figure*}

The radial distribution of baryonic components is plotted in
Fig.~\ref{fig:fbary(r)}, with the WMAP7 inferred Universal baryon
fraction shown for comparison. The gas fraction from the cluster model
is plotted as a solid curve with grey error envelope, with dots
indicating extrapolation beyond the data. Aperture values of the
stellar mass fraction (including the ICL) are plotted at \rtwofiveh\
and \rfiveh, with corresponding total baryon fraction (gas\,+\,stars)
plotted at the same radii. It can be seen that for all 5 clusters the
stellar fraction is higher within \rtwofiveh\ than \rfiveh, although
this difference is not statistically significant for A2401. This is
also consistent with the observation that the ICL in these systems is
more spatially concentrated than the galaxies themselves
\citepalias{gonzalez07}, as also found in the SDSS stacking analysis
of \citet{zibetti05}, for example. On the other hand, the gas fraction
profiles tend to increase monotonically with radius, with the
exception of AS84 and A2984, in accordance with the tendency for the
intracluster medium to be more extended than the dark matter
\citep[e.g.][]{david95,san03,vikhlinin06,sun09}. Interestingly, the
nearly flat \fgas\ profiles and more centrally-concentrated \fstar\
for AS84 and A2984 imply a \emph{decline} in total baryon content with
radius. However, this difference is not significant for AS84 and, for
A2984, \fbary\ within \rfiveh\ is only 1.7$\sigma$ lower than \fbary\
within \rtwofiveh.

The mass fractions for the baryon components averaged over all 5 poor
clusters are listed in Table~\ref{tab:baryfrac}, measured within
\rtwofiveh\ and \rfiveh\ and weighted by the inverse squared
measurement errors. The calculation was performed in log space since
the baryon fraction is right skewed, as a consequence of being
strictly bounded by zero below. This is confirmed by the fact that a
log-normal distribution provides a better description of the data than
a normal distribution \citep[as determined using the \textsc{fitdistr}
function in the \textsc{mass} package in \Rproject;][]{venables02}. It
can be seen that the combined baryon fraction within the two radii
does not significantly differ, reaching a value of $0.096\pm0.006$
within \rfiveh\ (at the corresponding mean \Mfiveh\ of
$(1.08^{+0.21}_{-0.18}) \times10^{14} M_{\sun}$), and falls some way
short (reaching only 57 per cent) of the Universal mean level of
$0.169\pm0.008$ from WMAP7 \citep{jarosik11}. On the other hand the
balance of hot and cold baryons does vary radially, with the ratio of
stellar to total baryon mass declining from 0.44 to 0.27 between
\rtwofiveh\ and \rfiveh.

Looking at Fig.~\ref{fig:fbary(r)}, if we consider only the systems
where the fitted data extend beyond \rfiveh\ (i.e. where no
extrapolation is required), it is clear that a substantial baryon
deficit is present in two of three cases, and is a factor of $\sim$2
for AS0296 (with a best-fitting value of $50\pm6$ per cent of the
Universal mean); for A2984, $\fbary=0.127\pm0.019$, representing
$75\pm11$ per cent of the Universal mean \fbary. We demonstrate in
Section~\ref{ssec:syserr-fbary} that, for 4/5 cases, the baryon
fractions cannot be reconciled with the Universal mean even when
conservatively accounting for systematic errors.

\begin{table}
\begin{center}
\begin{tabular}{lcc}
  \hline
 & $r_{\mathrm{2500}}$ & $r_{\mathrm{500}}$ \\ 
  \hline
$f_{\mathrm{star}}$ & 0.0450$_{-0.0067}^{+0.0078}$ & 0.0261$_{-0.0028}^{+0.0031}$ \\ 
  $f_{\mathrm{gas}}$ & 0.0580$_{-0.0105}^{+0.0128}$ & 0.0702$_{-0.0046}^{+0.0050}$ \\ 
  $f_{\mathrm{gas+stars}}$ & 0.1030$_{-0.0124}^{+0.0150}$ & 0.0963$_{-0.0054}^{+0.0058}$ \\ 
   \hline
M$_{\mathrm{tot}}$ ($10^{13}$ M$_{\sun}$) & 4.43$_{-0.94}^{+1.19}$ & 10.83$_{-1.76}^{+2.10}$ \\ 
   \hline
\end{tabular}
\caption{Weighted mean mass fractions for the baryon components and total mass, measured within \rfiveh\ and \rtwofiveh. The calculation was performed in log space, yielding an asymmetric standard error on the unlogged values.}
\label{tab:baryfrac}
\end{center}
\end{table}

\subsection{Variation in stellar and gas fraction}
\begin{figure*}
  \centering \subfigure{
    \includegraphics[width=9.25cm]{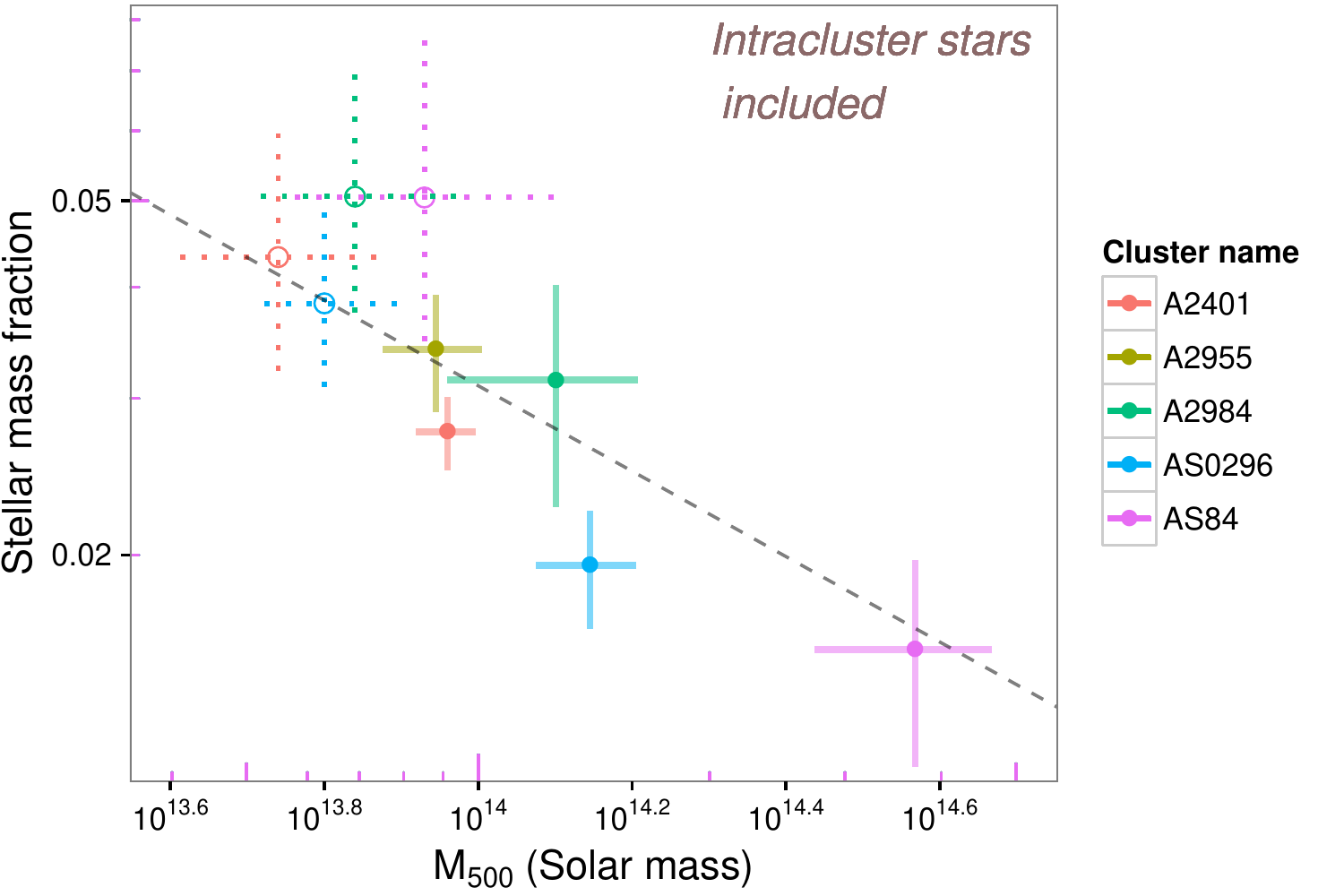}
  }
  \hspace{-2mm}
  \subfigure{
    \includegraphics[width=7.75cm]{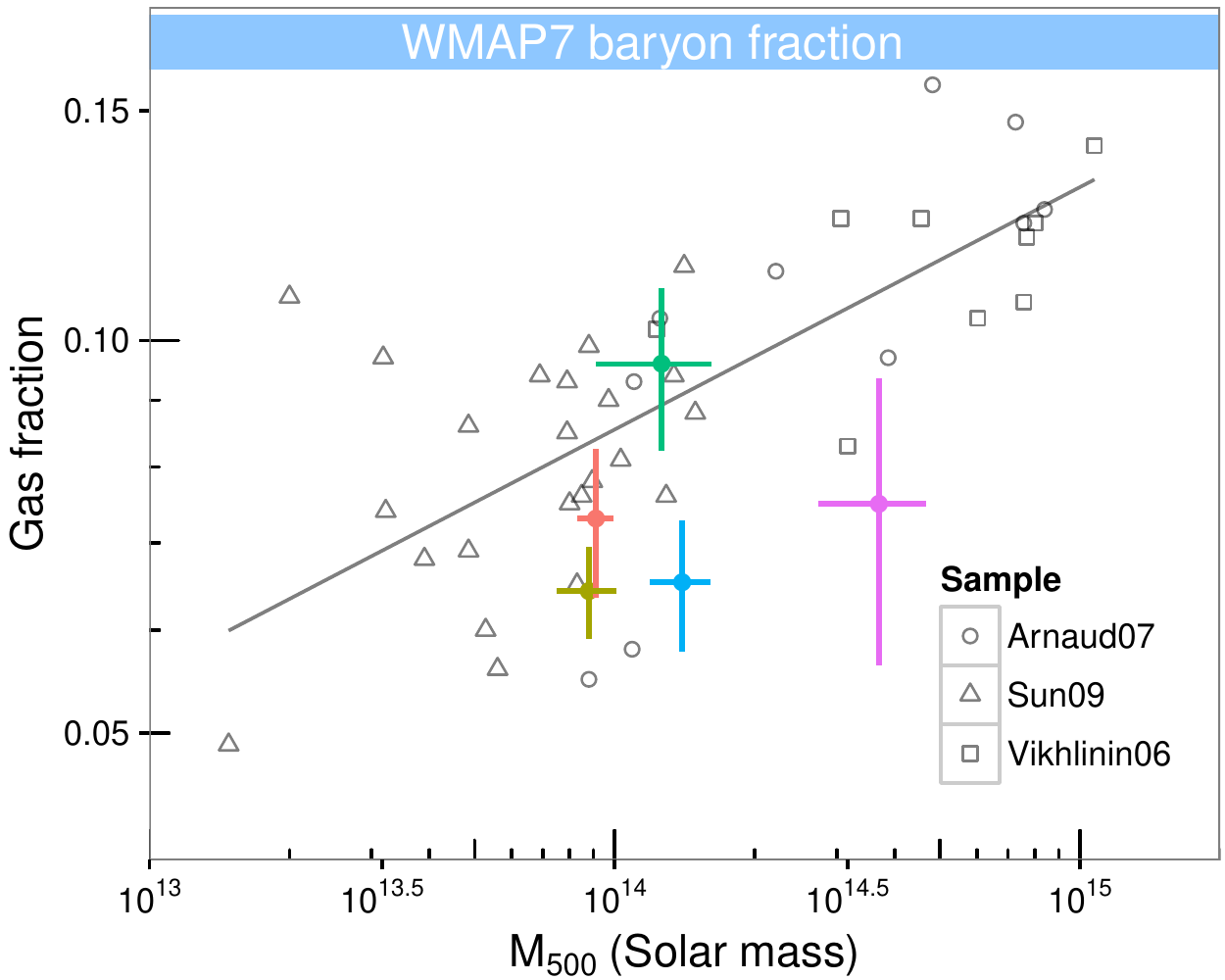}
  }
  \caption{\textit{Left:} Stellar mass fraction (including
    intracluster stars) vs. cluster total mass within \rfiveh,
    updating the results from \citetalias{gonzalez07} (dotted error
    bars, which were \emph{not} corrected for projection) with X-ray
    derived data from this work (solid error bars). The dashed line is
    the best fitting orthogonal regression from an analysis of new
    observations of the \citetalias{gonzalez07} sample (Gonzalez et
    al, in preparation). This sample includes the five clusters from
    this work, but mainly comprises more massive clusters. Note that
    A2955 was excluded from the original plot data of
    \citetalias{gonzalez07}, but would have been located off the left
    hand edge of the plot, and that an $I$ band mass-to-light ratio of
    2.65 is used to calculate stellar mass (see
    Section~\ref{ssec:ICL_Mstar} further details). \textit{Right:} gas
    fraction vs. mass for the same 5 clusters, together with X-ray
    selected data from the literature \citep{vikhlinin06, arnaud07,
      sun09}. The solid line is the best-fitting BCES power law
    regression to the X-ray selected literature data only
    (i.e. \emph{excluding} our 5 clusters), with logarithmic slope of
    $0.186\pm0.083$.}
  \label{fig:fstar/gas_vs_M500}
\end{figure*}

With our new X-ray derived total masses and the associated overdensity
radii, we are able to address one of the key results of
\citetalias{gonzalez07}, namely the stellar mass fraction in poor
clusters. The left panel of Fig.~\ref{fig:fstar/gas_vs_M500} shows
\fstar\ vs. $M_{500}$ for these 5 clusters. Also plotted, as dotted
error bars, are the positions of 4 of the clusters from the original
\citetalias{gonzalez07} analysis (the A2955 \fstar\ measurement was
excluded from that analysis, since its mass estimate was deemed
unreliable, being based on an extrapolation of the
\Mfiveh\,-\,$\sigma$ relation). It is clear that the new positions for
these points are substantially different, primarily due to the X-ray
masses being considerably larger than those derived from the velocity
dispersion, which we discuss further in Section~\ref{ssec:Mx_vs_Mopt}.

The larger masses result in a larger \rfiveh\ aperture within which
\fstar\ is measured and, since the BCG and ICL stellar contribution
falls off rapidly with radius (Section~\ref{ssec:radial_trends}), the
result is a \emph{lower} stellar fraction. However, it is interesting
that the change in both quantities acts to move the points mostly
along the regression line, thus largely preserving the trend found by
\citetalias{gonzalez07}.  While this is not altogether surprising
given the covariance between \fstar\ and \Mfiveh\ (since $\fstar
\propto \Mfiveh^{-1}$), the shift is nevertheless along a flatter
trajectory than implied by the relation between these quantities. For
reference, the dashed line in the left panel of
Fig.~\ref{fig:fstar/gas_vs_M500} shows the best fitting orthogonal
regression from an analysis of new observations of the (optically
selected) \citetalias{gonzalez07} sample (Gonzalez et al, in
preparation), which incorporates the results from this work. Despite
the substantial change in total mass and a reduction in the
measurement errors, the five clusters remain fairly consistent with
the trend and scatter in the \fstar\,--\,\Mfiveh\ relation from
\citetalias{gonzalez07}.

The corresponding relationship between gas fraction and mass is
plotted in the right panel of Fig.~\ref{fig:fstar/gas_vs_M500}. For
comparison, data from the X-ray selected cluster samples of
\citet{vikhlinin06} and \citet{arnaud07} as well as the X-ray selected
groups sample of \citet{sun09} are also shown, with error bars omitted
for clarity. The solid line is the best-fitting BCES orthogonal
regression \citep{akritas96} in log-log space, which excludes our 5
clusters and thus represents a purely X-ray selected sample. Of the 5
clusters, only A2984 is located near the trend, with the remaining 4
all lying near the lower edge of the scatter about the fit, in
contrast to the \fstar\,--\,\Mfiveh\ relation. This is not necessarily
surprising, since these were not X-ray selected clusters and therefore
might be expected to have lower \fgas\ (and hence \LX), as seen in
other optically-selected groups and clusters
\citep[e.g.][]{bower97b,rasmussen06b,rykoff08a,hicks08,dai10,balogh11}.
Indeed, \citet{rykoff08b} demonstrate that differences in the X-ray
luminosity--total mass relation between optically and X-ray selected
cluster samples can be at least partly explained by the combination of
Malmquist bias and deviations from hydrostatic equilibrium affecting
the latter.

\subsection{Variation in total baryon content}
\begin{figure}
\includegraphics[width=8.5cm]{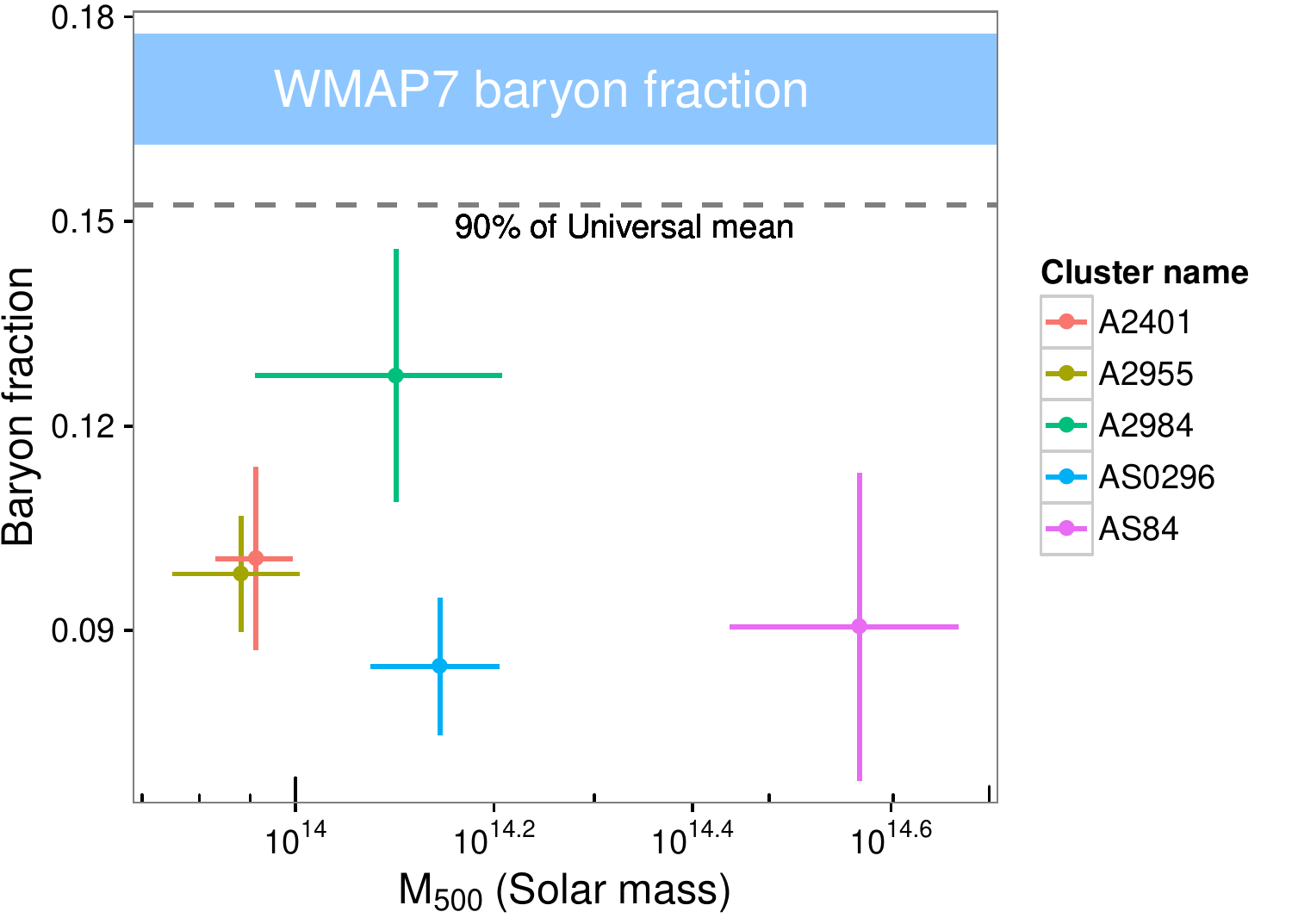}
\caption{ \label{fig:fbary-M500} Total baryon fraction (hot gas +
  stars) vs. total mass within \rfiveh\ for the 5 clusters. The
  horizontal dashed line shows the `depleted cosmic' \fbary\ typically
  expected within \rfiveh\ from simulations without radiative cooling
  or feedback \citep[e.g.][]{ettori06}. }
\end{figure}

The sums of the stellar (including ICL) and gas fractions within
\rfiveh\ for the five clusters are plotted in
Fig.~\ref{fig:fbary-M500}, to show the variation in total baryon
content compared to the Universal mean fraction. Fractional errors on
the total baryon fraction were calculated by adding in quadrature the
fractional errors on the gas and stellar fractions. Also indicated in
Fig.~\ref{fig:fbary-M500} is the level of a 10 per cent depleted
Universal \fbary\ (i.e. 0.9$\times\Omega_b / \Omega_m$ hereafter
referred to as the `depleted cosmic' \fbary), representing the typical
value seen within \rfiveh\ in simulated clusters
\citep[e.g.][]{ettori06}. It can be seen that 4 out of 5 of our
(optically-selected) clusters fall significantly ($> 2\sigma$) below
this line, implying a baryon deficit. With the exception of A2984,
this is mainly the result of a low gas fraction compared to the trend
(right panel Fig.~\ref{fig:fstar/gas_vs_M500}), although A2984 lies
close to the trend in both gas and stellar fraction but still shows
evidence of baryon shortfall. Furthermore, even when allowance is made
for systematic uncertainties, these baryon fractions cannot be
reconciled with the Universal mean (see
Section~\ref{ssec:syserr-fbary}).

\section{Discussion}
\label{sec:discuss}
An important aspect of our findings is that the X-ray determined
masses of the 5 clusters are all substantially greater than implied by
the velocity dispersion of their galaxy members. Therefore, in this
section we first address this issue, before turning to a discussion of
potential systematic effects relating to the gas and stellar mass
measurements and finally considering the interpretation of our results
and their associated implications.

\subsection{X-ray vs. optical mass estimates}
\label{ssec:Mx_vs_Mopt}
\begin{figure}
\includegraphics[width=8.5cm]{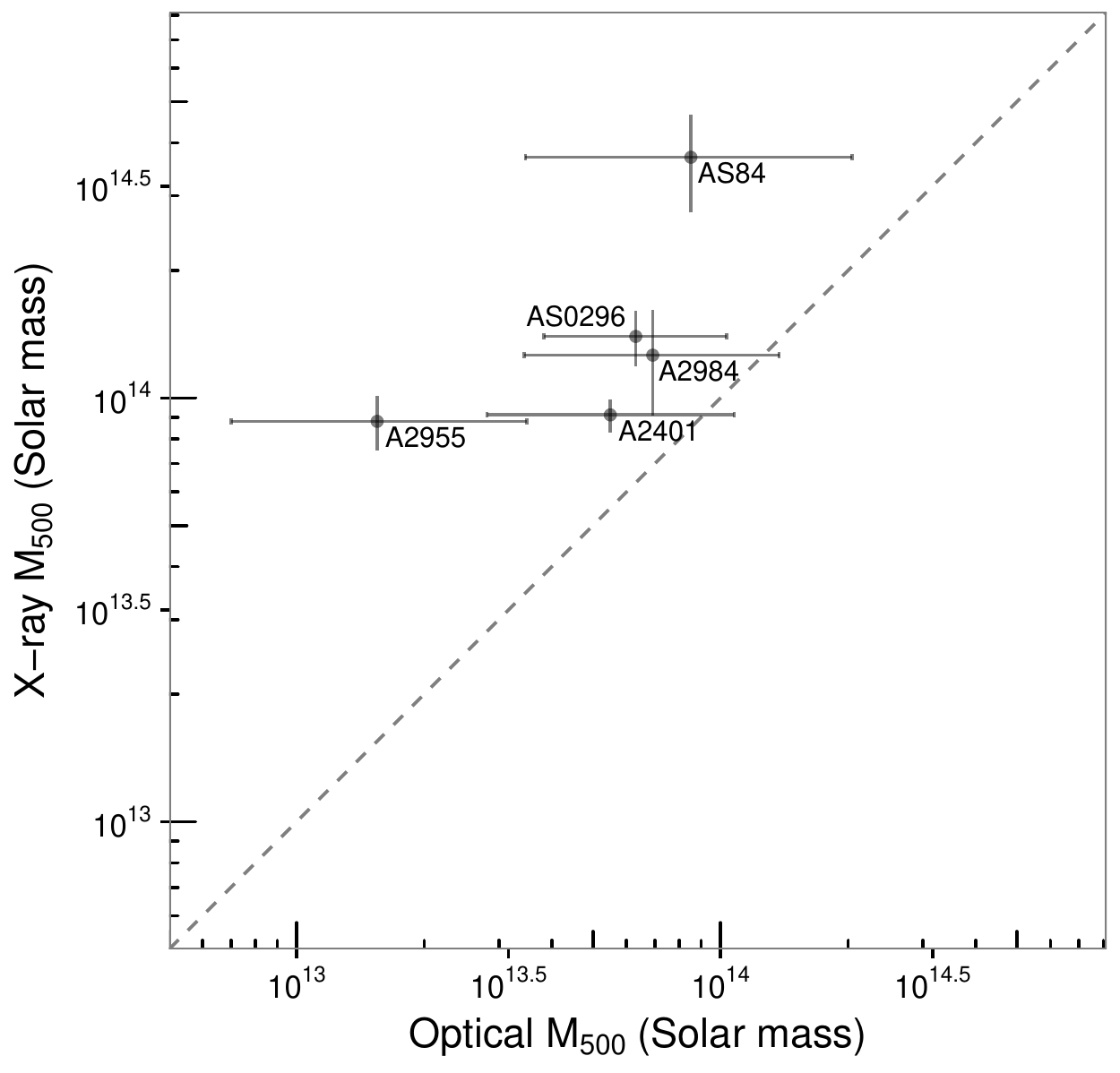}
\caption{ \label{fig:mass_comparison} A comparison of M$_{500}$ as
  measured from X-ray data in this work with the optically-derived
  estimates in \citetalias{gonzalez07}, which were based on the
  velocity dispersion of galaxies in the cluster and an assumed
  $\sigma$-\Mfiveh\ relation. The dashed line shows the locus of
  equality. }
\end{figure}

Estimating the gravitating mass of galaxy clusters is crucial in two
ways in determining their baryon fraction, since this enters as the
denominator in that quantity as well as determining the aperture
within which the measurement is made. The original
\citetalias{gonzalez07} analysis used the optical velocity dispersion
of the galaxies to estimate M$_{500}$ (via the \MT\ and
$\sigma-T_{\mathrm{X}}$ scaling relations) and this quantity is plotted
in Fig.~\ref{fig:mass_comparison} against the corresponding X-ray
inferred value measured in this work, with the locus of equality
indicated. In all cases the X-ray estimates are significantly higher
than their optical counterparts, by a factor of between 1.7 and 5.7
(mean 3.1).

Part of the discrepancy can be attributed to the calibration of the
$M-\sigma$ used by \citetalias{gonzalez07}, which was necessarily
limited by the lack of clusters with suitably well measured masses and
velocity dispersions. However, other potential causes of such a
discrepancy are failures of one or both of the two key assumptions in
both X-ray and optical mass estimates, namely equilibrium and
spherical geometry. On the X-ray side, non-thermal pressure support
and/or deviations from hydrostatic equilibrium (e.g. due to recent
disruption) are expected to cause X-ray mass \emph{under}estimates of
$\sim$5--20 per cent
\citep[e.g.][]{kay04,rasia06,nagai07,piffaretti08}; accounting for
this effect would only exacerbate the disagreement in this
case. Nevertheless, deviations from hydrostatic equilibrium are likely
to contribute towards the scatter in X-ray mass estimates, for which
we make allowance in our analysis of the systematic error budget in
Section~\ref{ssec:other_syserrs}.

With the exception of A2984 (where the uncertainties are large), none
of the clusters hosts a strong cool core (see
Table~\ref{tab:derivdata}), the absence of which is often associated
with recent disruption \citep[e.g.][]{san06}. On the other hand, there
is no indication of any significant offset between the X-ray
peak/centroid and the BCG, which suggests a lack of disturbance
\citep{san09b}. Moreover, since these are optically selected clusters,
it is plausible that they may have a lower incidence of cool cores
compared to X-ray selected samples, since the presence of a cool core
invariably boosts the X-ray flux in the region of highest surface
brightness.

Another possible explanation is that the line-of-sight velocity
dispersion is systematically underestimated for these clusters, which
could occur if the cluster was substantially elongated in the plane of
the sky. This hypothesis is difficult to test, given the limited
number of member galaxies, but the morphology of the isodensity
contours of all the surrounding galaxies (which includes fore- and
background objects as well as cluster members) hints at such a
possible elongation in the case of A2955 (Fig.~\ref{fig:radec-a2955})
and AS0296 (Fig.~\ref{fig:radec-aS0296}).  No such elongation is
evident for A2401 (Fig.~\ref{fig:radec-a2401}), and here the optical
and X-ray masses are in better agreement (although still significantly
discrepant). Interestingly, \citet{connelly12} have found that
dynamical masses have a large scatter and, for less massive clusters
in particular, are biased low compared to X-ray mass estimates. This
suggests that, for poor clusters in general, X-ray mass estimates may
be more reliable than those based on the velocity dispersion of the
member galaxies, particularly when the number of galaxies is
relatively small ($\sim$20--30; Table~\ref{tab:optdata}). Indeed,
\citet{biviano06} find that the scatter in $\sigma$-based mass
estimates roughly doubles when the number of galaxies decreases from
$\sim$400 to $\sim$20 (reaching $\sim$0.6 for \rmsub{N}{gal}=10),
based on an analysis of 62 simulated clusters. However, this alone is
not sufficient to account for the discrepancy in this case, which
amounts to a \emph{systematic bias} of at least a factor of $\sim$2.

\subsection{The impact of the total mass on \fbary}
\label{ssec:implied_fbary500}
The total gravitating mass determines the aperture for measuring the
gas and stellar masses, via the definition of an overdensity,
$\Delta$, with respect to the redshift-dependent critical density of
the Universe, $\rho_{\mathrm{crit}}(z)$
\begin{equation}
\label{eqn:overd}
 M_{\Delta} = \Delta \, \frac{4}{3} \pi \, r^3_{\Delta} \, \rho_{\mathrm{crit}}(z) \, .
\end{equation}
The stellar and gas fractions are inversely proportional to
$M_{\Delta}$, by definition, and in general both vary oppositely with
radius (Fig.~\ref{fig:fbary(r)}). As a result, \fbary\ within a given
overdensity radius, $r_{\Delta}$, depends quite subtly on the
corresponding $M_{\Delta}$, via Equation~\ref{eqn:overd}. Thus a large
change in \Mfiveh\ does not necessarily change the baryon fraction
within \rfiveh\ substantially. As a consequence, \fbary\ is quite
robust to variations in \Mfiveh\ resulting from modelling
uncertainties, for example.

\begin{figure*}
\includegraphics[width=17cm]{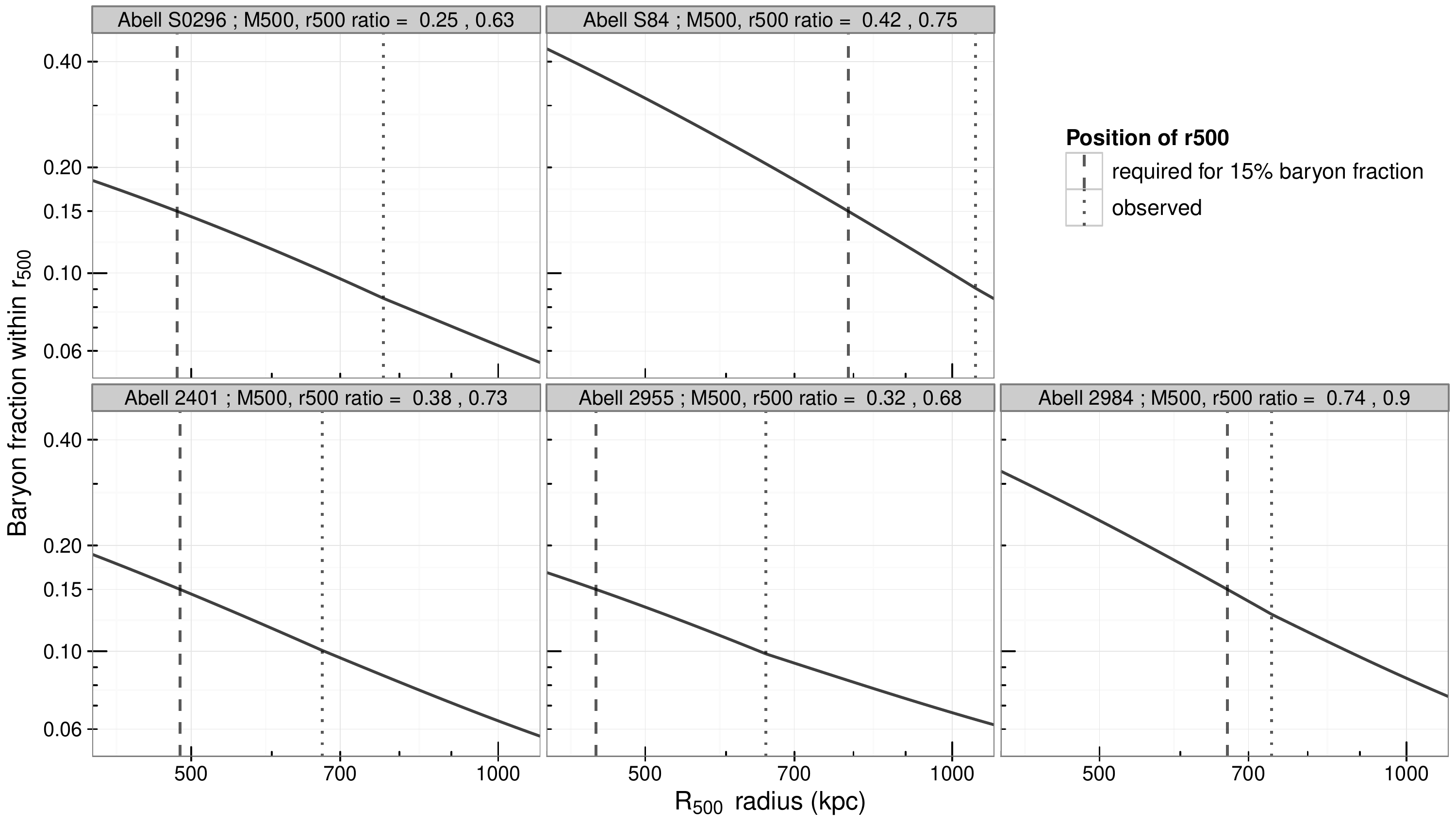}
\caption{ \label{fig:implied_fbaryprof} Baryon fraction profiles
  implied by assuming that each radius corresponds to \rfiveh, with
  dashed vertical lines marking the radius required to achieve a
  nominal `depleted cosmic' value of $\fbary=0.15$ within \rfiveh, as
  compared to the position of the observed value of \rfiveh\ (dotted
  lines). Ratios of the implied \Mfiveh\ and \rfiveh\ for the case of
  $\fbary = 0.15$ to the corresponding observed values are indicated
  in the strip text for each panel. See text for details.}
 \end{figure*}

The sensitivity of the baryon fraction measured within \rfiveh\ to the
assumed \Mfiveh\ can be judged by exploring the effect of artificially
altering the value of \rfiveh, as follows. The gas and stellar masses
are measured within each nominal value of \rfiveh\ and the
corresponding \fbary\ is calculated by adding them together and
dividing by the \Mfiveh\ calculated from Equation~\ref{eqn:overd}. The
stellar mass is calculated from \fstar, which is fixed at the spot
values for \rtwofiveh\ and \rfiveh\ for $r \le \rtwofiveh$ and $r \ge
\rfiveh$, respectively, and linearly interpolated in log-log space in
between. Fig.~\ref{fig:implied_fbaryprof} shows curves of the
resulting implied baryon fraction measured in this way within each
nominal \rfiveh\ aperture.

Under the hypothesis that the true \fbary\ is 15 per cent (roughly the
`depleted cosmic' \fbary\ level of 0.9 times the Universal mean), the
corresponding \rfiveh\ required to achieve this result is plotted as a
dashed vertical line in Fig.~\ref{fig:implied_fbaryprof}. For
comparison, the observed \rfiveh\ is shown as a dotted line and ratios
of the implied \Mfiveh\ and \rfiveh\ to their observed counterparts
are shown in the strip titles for each panel in the plot. With the
exception of A2984, the masses would need to have been overestimated
by $\sim$\,2--4 times to achieve $\fbary = 0.15$, equivalent to an
overestimate in mean temperature of $\sim$1.3--2.7 times. Thus the
observed deficiency of baryons can only be accounted for by a factor
of $\sim$\,2--4 systematic error in the determination of the total
mass. Such a large error greatly exceeds both the measurement errors
and our best estimate of the true systematic uncertainty, as we
demonstrate in Section~\ref{ssec:M500_check}.

\subsection{Cross-check of X-ray masses using the \MT\ relation}
\label{ssec:M500_check}
\begin{figure}
\includegraphics[width=8.5cm]{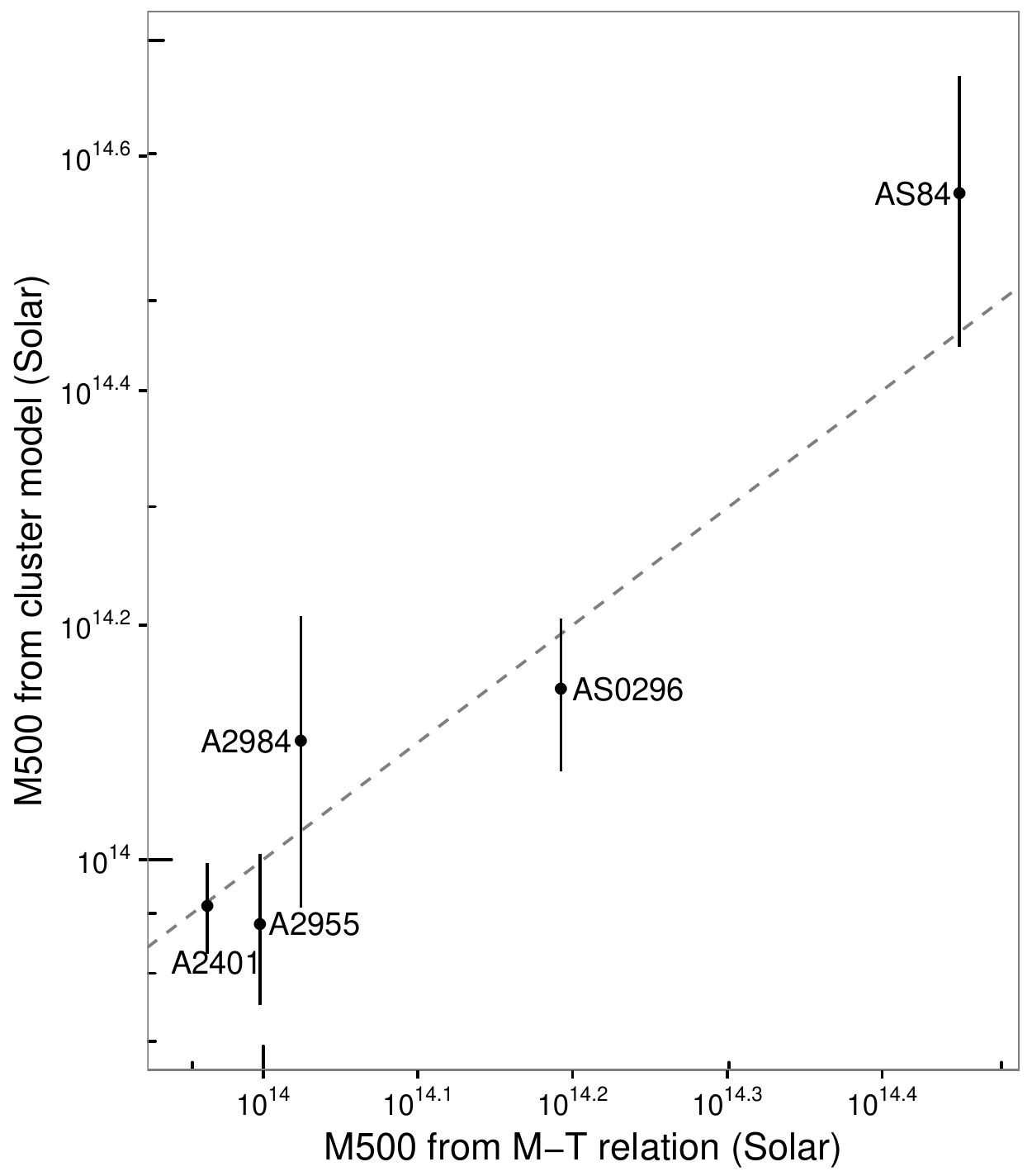}
\caption{ \label{fig:M-T_mass_comparison} A comparison of M$_{500}$
  determined from the cluster model with that estimated from the \MT\
  relation of \citet{vikhlinin06}. The dashed
  line shows the locus of equality. }
\end{figure}

To test the accuracy of the masses determined from the cluster
modelling, we also estimated \Mfiveh\ from the mass-temperature (\MT)
relation of \citet{vikhlinin06}, based on a global mean temperature
(\Tbar). We estimated \Tbar\ using \rhogasr\ and $T(r)$ from the
best-fitting cluster model, evaluating a weighted (by
$\rho^2_{\mathrm{gas}} \sqrt{T}$) mean in the radial range
(0.15--1)\,\rfiveh, determined iteratively (since \rfiveh\ depends on
\Mfiveh, which depends on \Tbar). The comparison between \Mfiveh\
calculated by these two methods is plotted in
Fig.~\ref{fig:M-T_mass_comparison}, which shows good agreement in all
cases and no indication of any systematic bias. This is consistent
with the good agreement in \rfiveh\ found by \citet{san10} between the
\citetalias{ascasibar08} cluster model and the
\citeauthor{vikhlinin06} \MT\ relation, using direct extraction and
fitting of spectra in an iteratively adjusted aperture
\citep[from][]{san09}. Although the \MT\ masses estimated in this way
are not completely independent from the \citetalias{ascasibar08} model
masses, this comparison nevertheless demonstrates consistency between
both the two methods and the cluster samples, in respect of
hydrostatic equilibrium. Since the \citeauthor{vikhlinin06} clusters
were carefully selected to have very regular X-ray morphology and with
only weak signs, if any, of dynamical activity, it is therefore
reasonable to infer that our cluster model-derived X-ray mass
estimates are reliable and not substantially affected by any
departures from hydrostatic equilibrium and/or spherical symmetry
within the hot gas haloes of the five clusters.

\begin{figure*}
\includegraphics[width=17cm]{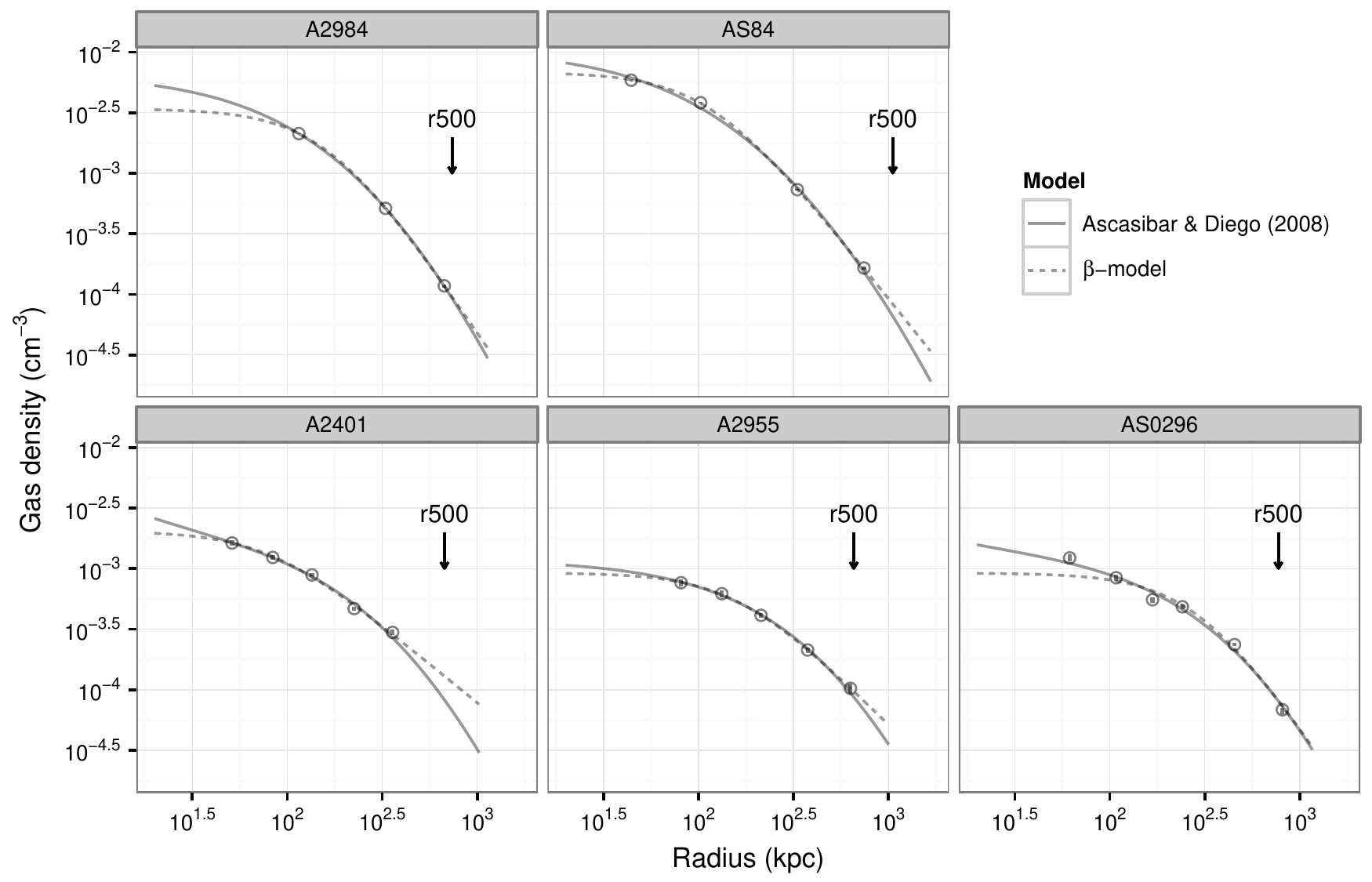}
\caption{ \label{fig:betamodels} A comparison of the best-fitting
  \citetalias{ascasibar08} and $\beta$-model gas density profiles,
  with the measured data plotted as hollow points and (mostly very
  small) error bars. Within each panel the profiles extend out to
  \rtwoh. For A2401, the excluded outermost two points (see
  Section~\ref{ssec:clusmodel}) are not shown, but both lie above the
  best-fit $\beta$-model.}
\end{figure*}

A further comparison of the total cluster mass is available for A2984
\& AS84 from the independent analysis of the same \XMM\ data by
\citet{sivanandam09}, also using the \citet{vikhlinin06} \MT\
relation, but measuring the mean temperature directly via spectral
fitting of the raw data. The values of \Mfiveh\ obtained by
\citeauthor{sivanandam09} are (in units of $10^{14}$\,M$_{\odot}$)
$0.96\pm0.098$ for A2984 (compared to $1.3\pm0.35$;
Table~\ref{tab:derivdata}) and $2.6\pm0.28$ for AS84 (compared to
$3.7\pm0.96$; Table~\ref{tab:derivdata}). In both cases our masses are
$\sim$1.4 times larger (albeit consistent to within the statistical
uncertainties), which places an upper bound on uncertainties
associated with the approach employed. As a result, we conclude that
our X-ray derived mass estimates (and corresponding \rfiveh\ and
\rtwofiveh\ apertures) are reliable and that their measurement errors
capture all reasonable sources of uncertainty. However, the impact of
the mass measurement method is incorporated into our estimate of the
systematic error on gas fraction at the end of
Section~\ref{ssec:betamodel}. The main reason for the systematic
difference in our masses compared to those of
\citeauthor{sivanandam09} is a difference in the steepness of the
outer gas density profile, which directly affects the pressure
gradient and hence the enclosed hydrostatic mass, and we investigate
this issue further in the next section.

\subsection{Choice of gas density model}
\label{ssec:betamodel}
It is noteworthy that the issue of the gas density in the outer
regions of galaxy groups and poor clusters is not well explored, but
it is clearly of importance in measuring the gas mass. It is now well
established that in more massive clusters the density profile beyond
$\ga$\,0.5\,\rfiveh\ typically steepens relative to a power law
extrapolated from smaller radii (i.e. a $\beta$-model), from studies
with \Rosat\ \citep{vikhlinin99}, \XMM\ \citep{croston08} and
\Chandra\ \citep[e.g.][]{ettori09}. However, no such steepening has
been found in galaxy groups, where a $\beta$-model provides a good
description of the density profile out to \rfiveh\
\citep{rasmussen04}. We therefore also fitted $\beta$-models to the
deprojected data points for all 5 clusters to cross-check the results
from the \citetalias{ascasibar08} cluster model, which does \emph{not}
use a $\beta$-model for the gas density profile.  The comparison is
made in Fig.~\ref{fig:betamodels} and it is clear that for A2955,
AS0296 \& A2984 there is excellent agreement between the two different
models. Moreover, in each of these cases the data extend beyond
\rfiveh, resulting in negligible differences in the gas mass within
\rfiveh\ ($\la$ 3 per cent) between the two models. Beyond this radius
the models tend to diverge, which reflects the lack of data points to
anchor the fit, while differences between the two models inside of the
innermost data point reflect the inability of the (flat cored)
$\beta$-model to capture the modest cuspiness in the density profiles.

In making this comparison, it should be noted firstly that the
\citetalias{ascasibar08} model is jointly fitted to both \rhogasr\ and
$T(r)$, assuming hydrostatic equilibrium and, also, that the
\citetalias{ascasibar08} model gas density profile is somewhat similar
to a $\beta$-model in the limit where there is no cool core component
(see figure~1 of \citetalias{ascasibar08})\,-- corresponding to $t =
1$, which is the case for A2955 (see Table~\ref{tab:jfitpars}
). Furthermore, the \citetalias{ascasibar08} model assumes a
polytropic ICM, which may not represent the true physical state of the
gas, although this assumption does not apply within the cool core (if
one is present-- i.e. $t<1$ in Table~\ref{tab:jfitpars}). In the case
of A2401 there is a difference between the extrapolated model profiles
beyond the outermost data point, with the \citetalias{ascasibar08}
model steepening compared to the $\beta$-model and yielding a lower
gas mass within \rfiveh\ by 15 per cent. On the other hand, the two
gas mass estimates agree to within 5 per cent for AS84, where
extrapolation is also needed to reach \rfiveh.

It is clear from the 3 clusters with the best radial data coverage
(A2984, A2955 \& AS0296; Fig.~\ref{fig:betamodels}) that no steepening
of the gas density profile logarithmic slope is observed, within at
least \rfiveh. This is quite different from the trend in more massive
clusters and we discuss the implications of this behaviour in
Section~\ref{ssec:feedback}. However, our results are in accordance
with \citet{rasmussen04}, who found that \rhogasr\ was well described
by a $\beta$-model over the entire radial range (which extended beyond
$\sim$\rfiveh) in two galaxy clusters of a similar mass scale
($\sim$2\,kev). On the other hand, \citet{sun09} claim that the
\citet{vikhlinin06} \rhogasr\ model (which features a steepening
logarithmic gradient) provides a good fit to their \Chandra\ sample of
43 galaxy groups (although a number of these systems are not well
covered azimuthally and/or radially beyond $\sim$0.5\rfiveh).

Finally, as was the case with the total mass, we are also able to
perform a cross-check of our gas mass results for A2985 and AS84, for
which the same \XMM\ data were analysed by \citet{sivanandam09}, using
a different analysis and modelling approach. They generated \XMM\ MOS
X-ray surface brightness images, with corresponding exposure maps and
background images and fitted these with a two-dimensional
$\beta$-model, convolved with the instrument point spread
function. They then deprojected this fitted profile using the Abel
transform, and applied a global spectral fit to determine the
corresponding normalization in terms of gas density. The values of
\Mgas\ ($<$\,\rfiveh) obtained by \citeauthor{sivanandam09} are (in
units of $10^{13}$\,M$_{\odot}$) $1.1\pm0.08$ for A2984 (compared to
$1.21\pm0.15$; based on the data shown in Table~\ref{tab:derivdata})
and $2.3\pm0.13$ for AS84 (compared to $2.77\pm0.61$;
Table~\ref{tab:derivdata}), which are not significantly different from
our own measurements.

Although our estimates of the separate systematic effects of total and
gas mass measurement method are not substantially different from zero,
they compound in the measurement of the gas fraction. Our best
estimate of the systematic error in measuring \fgas\ can be obtained
from a comparison of our results with those of \citet{sivanandam09},
who used different methods to us for determining both the total and
gas mass (albeit applied to the same \XMM\ data). The ratios of \fgas\
within \rfiveh\ measured with the \citetalias{ascasibar08} model
vs. \citeauthor{sivanandam09} method are $0.83\pm0.17$ and
$0.85\pm0.26$ (for A2984 \& AS84, respectively, with fractional errors
summed in quadrature). While neither of these ratios is significantly
different from 1, they are quite similar and somewhat less than
one. Taking our \citetalias{ascasibar08} model results as a baseline,
this represents a systematic increase of 15 per cent in \fgas\ as a
result of using a different analysis method. We therefore assign a
systematic uncertainty of +15 per cent associated with the
deprojection analysis and cluster modelling, but acting only in the
direction of increasing \fgas. This and the other contributions to the
systematic error budget for the gas fraction are summarised in the
middle column of Table~\ref{tab:syserror}, with negative values
quantifying any potential decrease in \fgas\ with respect to our
baseline model.

\subsection{Other systematic effects}
\label{ssec:other_syserrs}
The bootstrap re-sampling approach, used to evaluate uncertainties on
all quantities derived from the cluster model, yields a direct
estimate of the statistical error that takes full account of intrinsic
correlations (e.g between the gas and total mass in calculating \fgas)
and avoids the need for error propagation. However, it is important to
consider sources of systematic uncertainty that could bias our results
in a way that is not captured by these measurement errors.

As already mentioned in Section~\ref{ssec:clusmodel}, one source of
systematic error is gas clumping \citep[e.g.][and references
  therein]{urban11,nagai11}, which acts to bias the gas density
higher. Since it is more prevalent in cluster outskirts, clumping
would lead to a flatter density profile and a correspondingly
overestimated gas mass and \fgas, although this is difficult to
quantify. In a similar fashion, non-thermal pressure support or
merger-driven deviations from hydrostatic equilibrium could lead to an
overestimated gas fraction, by $\sim$5--20 per cent\,-- the amount by
which the total mass is likely underestimated
\citep[e.g.][]{kay04,rasia06,nagai07,piffaretti08}. In any case, both
these phenomena would lead to problems in fitting the
\citetalias{ascasibar08} cluster model and likely result in a
discrepant gas mass compared to that estimated from a surface
brightness deprojection, using a $\beta$-model for example. While we
cannot rule this out, the indications from
Section~\ref{ssec:betamodel} are that any such differences are small
compared to our measurement errors. 

To calculate the impact on \fgas\ and \fstar\ of a +20 per cent
systematic error on \Mfiveh\ resulting from non-hydrostatic bias, we
adopt a similar approach to that detailed in
Section~\ref{ssec:implied_fbary500}. A new \rfiveh\ is calculated
based on $1.2\times\Mfiveh$ (via Equation~\ref{eqn:overd}) and this
aperture is used to measure the gas and stellar mass, which are then
divided by $1.2\times\Mfiveh$. In this case the stellar mass is
calculated by simple extrapolation of the linear interpolation in
log-log space of the spot values of \fstar\ measured at \rtwofiveh\
and \rfiveh. The resulting changes for the 5 clusters have mean values
and standard errors of -($10\pm0.6$) per cent for \fgas\ and
-($4\pm0.6$) per cent for \fstar. Since this bias only acts one way
(to cause the X-ray inferred mass to be underestimated), we take these
mean values as the negative systematic uncertainty on these
quantities, with a corresponding zero contribution in the positive
direction (see Table~\ref{tab:syserror}).

%
\begin{table}
\centering
\begin{tabular}{r*{2}{c}}
\hline \\[-2ex]
  & \multicolumn{2}{c}{Percentage contribution} \\
 Systematic error source & \fgas\ & \fstar\ \\
\hline\\[-2ex] 
Cluster modelling & +15 / -0 & -- \\
Non-hydrostatic equilibrium & +0 / -10 & +0 / -4 \\
Outer bin density correction factor & $\pm7$ & -- \\
\XMM\ calibration & $\pm5$ & -- \\
Luminosity function & -- & $\pm8$ \\
Mass to light ratio & -- & $\pm15$ \\
\hline
Total & +27 / -22 & +23 / -28 \\
\hline
\end{tabular}
\caption{Estimates of the positive and negative percentage contributions 
 to the systematic errors on the gas and stellar fractions.}
\label{tab:syserror}
\end{table}

Another source of systematic error is the model-dependent correction
factor used to determine the gas density in the outermost bin
(Section~\ref{ssec:clusmodel}). In the case of A2955 \& AS0296, we
find that refitting the cluster model with this outermost bin omitted
leads to a gas fraction within \rfiveh\ which is 6 per cent lower and
19 per cent higher, respectively. This is comparable with the
fractional measurement errors (of 8 \& 12 per cent) and, in view of
the inconsistency of direction, we have assigned an average
(symmetric) systematic error of $\pm7$ per cent to account for this
effect (i.e. the mean of -6 and +19, rounded up; see
Table~\ref{tab:syserror}).

Since our X-ray analysis is based entirely on \XMM\ observations, we
also need to consider the possible impact of systematic calibration
uncertainties. Although this is difficult to quantify, the most recent
(Dec 2010) results of \XMM\ cross-calibration with \Chandra\
indicate\footnote{\url{http:/xmm2.esac.esa.int/external/xmm_sw_cal/calib/documentation/index.shtml/}}
that the effective area is subject to a $\sim$5--10 per cent
uncertainty, which would translate into a 2.5--5 per cent systematic
uncertainty on the gas density/mass. However, this is based on older
versions of the analysis software and the calibration of both \XMM\
and \Chandra\ has improved since then. Nevertheless, we conservatively
assume a (symmetric) systematic error of $\pm5$ per cent resulting
from this (see Table~\ref{tab:syserror}).

Sources of systematic error on the stellar fraction are somewhat
harder to quantify, but are dominated by the choice of mass-to-light
ratio used to convert luminosity to stellar mass. As previously
mentioned, our chosen value of $I$ band $M/L$ = 2.65 is somewhat lower
than the value of 3.6 used originally by \citetalias{gonzalez07}, but
which now corrects for a 15 per cent dark matter contribution to the
mass. The two main sources of systematic error in $M/L$ are the dark
matter contribution and the faint end slope of the luminosity
function, which we estimate to be $\pm15$ and $\pm8$ per cent,
respectively (rightmost column of Table~\ref{tab:syserror}; Gonzalez
et al, in preparation).

Finally, it is worth noting that the impact of selection effects may
be significant here, since the \citetalias{gonzalez07} parent sample
was selected on the basis of cluster optical properties and almost all
large studies of cluster gas mass/fraction to date use X-ray
selection. Selection effects are likely to lead to systematic
differences in gas fraction, which will be biased higher for X-ray
selected clusters; optically selected clusters could be biased low in
\fgas, if uncollapsed groups are included, although that is not the
case for the five clusters considered here. On the basis of the
weighted averages quoted in Table~\ref{tab:baryfrac}, the gas fraction
for our (optically-selected) sample is $\sim$$(19\pm6)$ per cent lower
than that predicted from the BCES regression to the X-ray selected
literature sample at our corresponding mean \Mfiveh. While
optically-selected clusters (and particularly groups) often have at
least a modest incidence of non-detections
\citep[e.g.][]{rasmussen06b,balogh11}, it is interesting that the
\XMM\ follow-up of the \citetalias{gonzalez07} sample has yielded good
X-ray detections in all cases, with sufficient data quality to enable
the gas fraction to be measured. However, we have ignored any
systematic uncertainty associated with selection effects, since these
are not relevant to the issue of baryon deficiency in individual
clusters.

\subsection{Systematic error on the baryon fraction}
\label{ssec:syserr-fbary}
To assess the impact of systematic errors on the total baryon fraction
within \rfiveh, we consider the separate contributions from the gas
and stellar fractions.  Based on our accounting of the sources of
systematic uncertainty listed in Table~\ref{tab:syserror}, the most
conservative aggregate values obtained (by a separate summation of the
individual positive and negative contributions) are +27/-22 per cent
for \fgas\ and +23/-28 per cent for \fstar. This assumes the most
extreme case where contributions with opposite signs (e.g. the cluster
modelling and non-hydrostatic biases) do \emph{not} cancel each other
out.

\begin{figure}
\includegraphics[width=8.5cm]{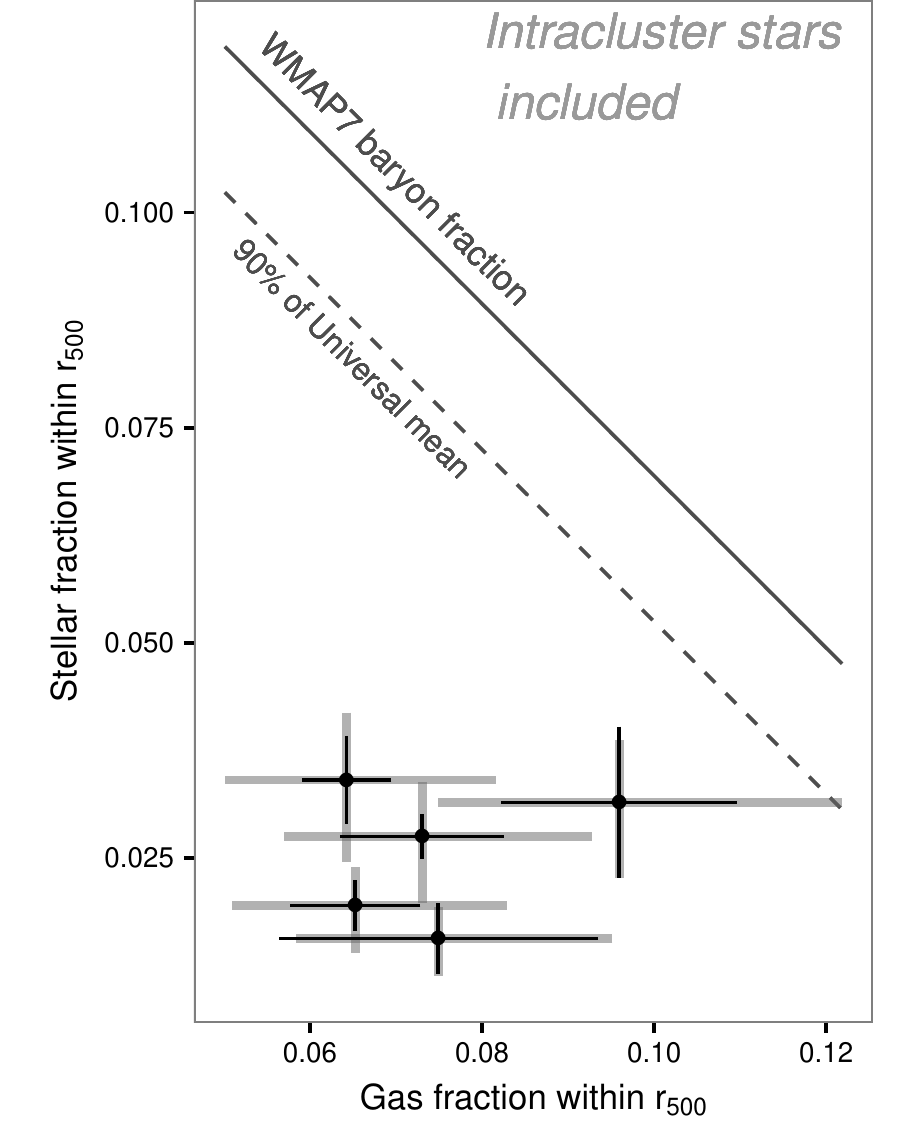}
\caption{ \label{fig:syserror-fbary} The location of the five clusters
  in the \fstar-\fgas\ plane, compared to the Universal mean baryon
  fraction. The two different types of error bar show the separate
  contributions from 1$\sigma$ measurement (thin black) and
  conservative systematic (thick grey) uncertainties (see
  Table~\ref{tab:syserror}).}
\end{figure}

The relative importance of both systematic and measurement errors on
the measurement of the total baryon fraction can be gauged from
Fig.~\ref{fig:syserror-fbary}, where the \fstar\ is plotted against
\fgas\ (both measured within \rfiveh). The measurement errors are
shown as thin black lines, with systematic errors plotted in thick
grey bars. The Universal mean \fbary\ and `depleted cosmic' \fbary\
lie along diagonal lines in this plane, oriented at 45\degr, as a
result of using an equal scale per unit length on each axis. Although
the (conservative) systematics dominate the measurement uncertainties
for most of these systems, it is clear that at least 4 of the 5
clusters lie well below these limits. We can therefore be confident
that a significant baryon deficit is present in these cases.

\subsection{Feedback and missing baryons}
\label{ssec:feedback}
There is considerable variation in \fbary\ for these five clusters,
which results mainly from the scatter in \fgas\
(Fig.~\ref{fig:fstar/gas_vs_M500}, right panel), as the dominant
baryonic component. While there are no indications of a systematic
variation of \fbary\ with total mass (Fig.~\ref{fig:fbary-M500}), the
baryon fractions of these five clusters are lower than those of
Gonzalez et al. (in preparation), which are generally more massive.
Moreover, in the case of the 3 clusters where we can directly measure
both gas and stars out to \rfiveh, there is evidently substantial
baryon loss. However, it is not clear if this points to a trend with
mass, or is the result of a break in self-similarity, as seen to occur
at $kT \approx 3.5$\,keV (corresponding to
$\sim2.6\times10^{14}\Msol$) in a recent analysis of the X-ray
luminosity-temperature relation \citep{maughan12}.

It is well established that cosmological simulations without radiative
cooling or galaxy feedback produce clusters that have baryon fractions
within \rfiveh\ that are consistent with the `depleted cosmic' \fbary\
level of $\sim$0.9 \citep[e.g.][]{ettori06}.  The introduction of
cooling allows baryons to transfer from the hot to the cold phase to
form stars and leads to a systematic variation in gas fraction with
halo mass \citep[e.g.][]{muanwong01}. Indeed, \citet{bode08} point out
that if stellar fraction varies as steeply as \Mfiveh$^{-0.49}$, then
cooling alone can explain the trend of gas fraction with halo
mass. However, radiative cooling produces no net loss of baryons--
this is clearly inconsistent with our results, which also include the
contribution from intracluster stars.  Furthermore, the
\citeauthor{bode08} model assumes that the stars have the same radial
distribution as the dark matter, whereas our results and those of
\citetalias{gonzalez07} favour a less extended light distribution than
that of the dark matter.

One possible explanation for the missing baryons is therefore galaxy
feedback, for which the phenomenon of AGN outflows is the most likely
candidate \citep[e.g.][]{short09,mccarthy10,mccarthy11}, whether prior
to or after gravitational collapse.  Indeed, it is now known that the
feedback associated with galaxy formation and black hole growth is
capable of altering the distribution of both baryonic \emph{and} dark
matter even on large scales, thereby rendering cosmological inference
sensitive to this localized \emph{baryonic} physics
\citep{lueker10,vanDaalen11}. It is therefore of great importance to
understand how cosmic feedback operates, and the corresponding balance
of hot and cold baryons in groups and clusters of galaxies. Under this
scenario, a diversity in feedback histories amongst the clusters could
also account for the scatter in \fgas, which dominates the variation
in the total baryon content.

Furthermore, the phenomenon of power-law-like \rhogasr\ in the
outskirts of groups compared to a steepening in clusters is also seen
in recent cosmological simulations where feedback has been
incorporated and found able and necessary to produce galaxy groups
that match many of the characteristics of observed systems
\citep{mccarthy10}. For example, the shape of the median
(group-dominated) gas entropy profile in figure~1 of
\citet{mccarthy10} roughly comprises a steeper inner power law
($<0.2\rfiveh$), turning over to a flatter slope until
$\sim$0.9\,\rfiveh, before steepening again. This behaviour is a
consequence of the increased impact of galaxy feedback in the
shallower potential wells of groups compared to clusters which,
firstly, displaces gas from inner to outer regions and, secondly,
lowers the ICM pressure and therefore causes more gas accretion from
outside the group (I. McCarthy, private communication). The net effect
is a flattening of \rhogasr\ around \rfiveh\ in groups compared to
clusters.

\section{Conclusions}
We have analysed the hot gas and stellar baryon content of five
low-mass galaxy clusters drawn from the optically-selected sample of
\citetalias{gonzalez07}, which includes both a direct mapping of the
intracluster light (ICL) as well as the galaxies' contribution, fully
corrected for the effects of projection. We use hydrostatic X-ray
cluster modelling of \XMM\ observations to measure the gas total mass
of these objects and the corresponding overdensity radii within which
to perform an inventory of baryons (\rtwofiveh\ and \rfiveh). We
summarize our main findings as follows.

\begin{enumerate}\setlength{\itemsep}{1ex}
\item For all five clusters our hydrostatic X-ray masses are well in
  excess of those originally estimated by \citetalias{gonzalez07} on
  the basis of their galaxy velocity dispersions ($\sigma$) and an
  assumed $\sigma-\Mfiveh$ relation, by a factor of between 1.7 and
  5.7 times. We also calculate alternative estimates of these X-ray
  masses using the \MT\ relation of \citet{vikhlinin06}, and consider
  a range of possible explanations for this discrepancy
  (Sections~\ref{ssec:Mx_vs_Mopt} \& \ref{ssec:M500_check}).

\item We study the stellar fraction vs. \Mfiveh, compared to the other
  clusters with similar ICL mapping in \citetalias{gonzalez07} and
  find that, despite the large change in total mass, the 5 clusters
  largely shift along the trend (since the higher mass implies larger
  \rfiveh, within which \fstar\ is measured). 

\item Within \rfiveh, we find weighted mean stellar (including ICL),
  gas and total baryon mass fractions of $0.026\pm0.003$,
  $0.070\pm0.005$ and $0.096\pm0.006$, respectively, at a
  corresponding weighted mean \Mfiveh\ of ($1.08_{-0.18}^{+0.21})
  \times 10^{14}$\,M$_{\odot}$ (see Table~\ref{tab:baryfrac}). This
  baryon fraction amounts to 57 per cent of the Universal mean, and
  implies a deficit of at least $\sim$30 per cent compared to the
  `depleted cosmic' \fbary\ level of
  $\sim$0.9$\times\Omega_b/\Omega_m$ expected in the absence of
  cooling and feedback \citep[e.g.][]{ettori06}. We consider the main
  sources of systematic error affecting our results and conclude that
  these are not sufficient to explain the observed shortfall in
  baryons in 4 of the 5 clusters.

\item In the three cases where we can trace the X-ray emission out to
  \rfiveh, we find no evidence for steepening of the gas density
  profile, as commonly seen in more massive clusters. This behaviour
  in consistent with the increased effect of galaxy feedback in
  shallower potential wells seen in the OWLS cosmological simulations,
  which leads to gas accumulation around \rfiveh\ (see
  Section~\ref{ssec:feedback}). This may also account for the baryon
  deficit observed within \rfiveh, and suggests that the missing
  baryons are located beyond this radius.
  
\end{enumerate}

As a result of the upward revision in mass, these systems are now no
longer located as far towards the extreme low-mass end of the trend as
previously thought \citepalias{gonzalez07}, at a scale where the
impact of cooling and feedback are most pronounced. Indeed, combined
ICL photometry and X-ray mapping of many more low mass groups is
clearly needed, to constrain the baryon balance in this important
regime. However, this remains observationally challenging, owing to
the tension between the need for proximity ($z \ll 0.1$) in X-ray
detection of faint group emission vs. the requirement of more compact
angular size to provide stable and well-characterized backgrounds for
diffuse optical light mapping, which is better achieved nearer $z \sim
0.1$. Given the uncertainly in modelling the gas density profile in
the outskirts of groups and low-mass clusters, it is also very
important to secure deep X-ray observations that can reach beyond
\rfiveh\ at high signal-to-noise ratio, to avoid extrapolation bias
and thereby measure the hot gas mass directly. This would have the
added benefit of probing a mass scale and spatial regions which are
sensitive to cosmic feedback, and may help uncover at least some of the
missing baryons.

\section*{Acknowledgments}
AJRS acknowledges support from the Science and Technology Facilities
Council (STFC) and EOS acknowledges support from an EU Marie Curie
fellowship.  This work made use of the NASA/IPAC Extragalactic
Database (NED) and data from The Two Micron All Sky Survey
(2MASS). AJRS thanks Ian McCarthy for useful discussions, and
acknowledges the excellent \textsc{ggplot2} package in
\Rproject\ \citep{wickham09} which was used to create the plots in
this paper.

\bibliography{/data/ajrs/stuff/latex/ajrs_bibtex}

\begin{thebibliography}{67}
\expandafter\ifx\csname natexlab\endcsname\relax\def\natexlab#1{#1}\fi

\bibitem[{{Akritas} \& {Bershady}(1996)}]{akritas96}
{Akritas}, M.~G., \& {Bershady}, M.~A. 1996, ApJ, 470, 706

\bibitem[{{Arnaud} {et~al.}(2007){Arnaud}, {Pointecouteau}, \&
  {Pratt}}]{arnaud07}
{Arnaud}, M., {Pointecouteau}, E., \& {Pratt}, G.~W. 2007, A\&A, 474, L37

\bibitem[{{Ascasibar} \& {Diego}(2008)}]{ascasibar08}
{Ascasibar}, Y., \& {Diego}, J.~M. 2008, MNRAS, 383, 369

\bibitem[{{Balogh} {et~al.}(2011){Balogh}, {Mazzotta}, {Bower}, {Eke},
  {Bourdin}, {Lu}, \& {Theuns}}]{balogh11}
{Balogh}, M.~L., {Mazzotta}, P., {Bower}, R.~G., {et~al.} 2011, MNRAS, 412, 947

\bibitem[{{Balogh} {et~al.}(2008){Balogh}, {McCarthy}, {Bower}, \&
  {Eke}}]{balogh08}
{Balogh}, M.~L., {McCarthy}, I.~G., {Bower}, R.~G., \& {Eke}, V.~R. 2008,
  MNRAS, 385, 1003

\bibitem[{{Beers} {et~al.}(1990){Beers}, {Flynn}, \& {Gebhardt}}]{beers90}
{Beers}, T.~C., {Flynn}, K., \& {Gebhardt}, K. 1990, AJ, 100, 32

\bibitem[{{Biviano} {et~al.}(2006){Biviano}, {Murante}, {Borgani}, {Diaferio},
  {Dolag}, \& {Girardi}}]{biviano06}
{Biviano}, A., {Murante}, G., {Borgani}, S., {et~al.} 2006, A\&A, 456, 23

\bibitem[{{Bode} {et~al.}(2009){Bode}, {Ostriker}, \& {Vikhlinin}}]{bode08}
{Bode}, P., {Ostriker}, J.~P., \& {Vikhlinin}, A. 2009, ApJ, 700, 989

\bibitem[{{Bower} {et~al.}(1997){Bower}, {Castander}, {Ellis}, {Couch}, \&
  {Boehringer}}]{bower97b}
{Bower}, R.~G., {Castander}, F.~J., {Ellis}, R.~S., {Couch}, W.~J., \&
  {Boehringer}, H. 1997, MNRAS, 291, 353

\bibitem[{{Connelly} {et~al.}(2012){Connelly}, {Wilman}, {Finoguenov}, {Hou},
  {Mulchaey}, {McGee}, {Balogh}, {Parker}, {Saglia}, {Henderson}, \&
  {Bower}}]{connelly12}
{Connelly}, J.~L., {Wilman}, D.~J., {Finoguenov}, A., {et~al.} 2012, ApJ, 756,
  139

\bibitem[{{Croston} {et~al.}(2008){Croston}, {Pratt}, {B{\"o}hringer},
  {Arnaud}, {Pointecouteau}, {Ponman}, {Sanderson}, {Temple}, {Bower}, \&
  {Donahue}}]{croston08}
{Croston}, J.~H., {Pratt}, G.~W., {B{\"o}hringer}, H., {et~al.} 2008, A\&A,
  487, 431

\bibitem[{{Dai} {et~al.}(2010){Dai}, {Bregman}, {Kochanek}, \& {Rasia}}]{dai10}
{Dai}, X., {Bregman}, J.~N., {Kochanek}, C.~S., \& {Rasia}, E. 2010, ApJ, 719,
  119

\bibitem[{{David} {et~al.}(1995){David}, {Jones}, \& {Forman}}]{david95}
{David}, L.~P., {Jones}, C., \& {Forman}, W. 1995, ApJ, 445, 578

\bibitem[{{Eckert} {et~al.}(2012){Eckert}, {Vazza}, {Ettori}, {Molendi},
  {Nagai}, {Lau}, {Roncarelli}, {Rossetti}, {Snowden}, \&
  {Gastaldello}}]{eckert12}
{Eckert}, D., {Vazza}, F., {Ettori}, S., {et~al.} 2012, A\&A, 541, A57

\bibitem[{{Ettori} \& {Balestra}(2009)}]{ettori09}
{Ettori}, S., \& {Balestra}, I. 2009, A\&A, 496, 343

\bibitem[{{Ettori} {et~al.}(2006){Ettori}, {Dolag}, {Borgani}, \&
  {Murante}}]{ettori06}
{Ettori}, S., {Dolag}, K., {Borgani}, S., \& {Murante}, G. 2006, MNRAS, 365,
  1021

\bibitem[{{Feldmeier} {et~al.}(2002){Feldmeier}, {Mihos}, {Morrison}, {Rodney},
  \& {Harding}}]{feldmeier02}
{Feldmeier}, J.~J., {Mihos}, J.~C., {Morrison}, H.~L., {Rodney}, S.~A., \&
  {Harding}, P. 2002, ApJ, 575, 779

\bibitem[{{Gonzalez} {et~al.}(2005){Gonzalez}, {Zabludoff}, \&
  {Zaritsky}}]{gonzalez05}
{Gonzalez}, A.~H., {Zabludoff}, A.~I., \& {Zaritsky}, D. 2005, ApJ, 618, 195

\bibitem[{{Gonzalez} {et~al.}(2007){Gonzalez}, {Zaritsky}, \&
  {Zabludoff}}]{gonzalez07}
{Gonzalez}, A.~H., {Zaritsky}, D., \& {Zabludoff}, A.~I. 2007, ApJ, 666, 147

\bibitem[{{Grevesse} \& {Sauval}(1998)}]{grevesse98}
{Grevesse}, N., \& {Sauval}, A.~J. 1998, Space Science Reviews, 85, 161

\bibitem[{{Hernquist}(1990)}]{hernquist90}
{Hernquist}, L. 1990, ApJ, 356, 359

\bibitem[{{Hicks} {et~al.}(2008){Hicks}, {Ellingson}, {Bautz}, {Cain},
  {Gilbank}, {Gladders}, {Hoekstra}, {Yee}, \& {Garmire}}]{hicks08}
{Hicks}, A.~K., {Ellingson}, E., {Bautz}, M., {et~al.} 2008, ApJ, 680, 1022

\bibitem[{{Hou} {et~al.}(2009){Hou}, {Parker}, {Harris}, \& {Wilman}}]{hou09}
{Hou}, A., {Parker}, L.~C., {Harris}, W.~E., \& {Wilman}, D.~J. 2009, ApJ, 702,
  1199

\bibitem[{{Jansen} {et~al.}(2001){Jansen}, {Lumb}, {Altieri}, {Clavel}, {Ehle},
  {Erd}, {Gabriel}, {Guainazzi}, {Gondoin}, {Much}, {Munoz}, {Santos},
  {Schartel}, {Texier}, \& {Vacanti}}]{jansen01}
{Jansen}, F., {Lumb}, D., {Altieri}, B., {et~al.} 2001, A\&A, 365, L1

\bibitem[{{Jarosik} {et~al.}(2011)}]{jarosik11}
{Jarosik}, N., {et~al.} 2011, ApJS, 192, 14

\bibitem[{{Kalberla} {et~al.}(2005){Kalberla}, {Burton}, {Hartmann}, {Arnal},
  {Bajaja}, {Morras}, \& {P{\"o}ppel}}]{kalberla05}
{Kalberla}, P.~M.~W., {Burton}, W.~B., {Hartmann}, D., {et~al.} 2005, A\&A,
  440, 775

\bibitem[{{Kay} {et~al.}(2004){Kay}, {Thomas}, {Jenkins}, \& {Pearce}}]{kay04}
{Kay}, S.~T., {Thomas}, P.~A., {Jenkins}, A., \& {Pearce}, F.~R. 2004, MNRAS,
  355, 1091

\bibitem[{{Krick} \& {Bernstein}(2007)}]{krick07}
{Krick}, J.~E., \& {Bernstein}, R.~A. 2007, AJ, 134, 466

\bibitem[{{Lin} {et~al.}(2004){Lin}, {Mohr}, \& {Stanford}}]{lin04}
{Lin}, Y., {Mohr}, J.~J., \& {Stanford}, S.~A. 2004, ApJ, 610, 745

\bibitem[{{Lueker} {et~al.}(2010)}]{lueker10}
{Lueker}, M., {et~al.} 2010, ApJ, 719, 1045

\bibitem[{{Maughan} {et~al.}(2012){Maughan}, {Giles}, {Randall}, {Jones}, \&
  {Forman}}]{maughan12}
{Maughan}, B.~J., {Giles}, P.~A., {Randall}, S.~W., {Jones}, C., \& {Forman},
  W.~R. 2012, MNRAS, 421, 1583

\bibitem[{{McCarthy} {et~al.}(2011){McCarthy}, {Schaye}, {Bower}, {Ponman},
  {Booth}, {Dalla Vecchia}, \& {Springel}}]{mccarthy11}
{McCarthy}, I.~G., {Schaye}, J., {Bower}, R.~G., {et~al.} 2011, MNRAS, 412,
  1965

\bibitem[{{McCarthy} {et~al.}(2010){McCarthy}, {Schaye}, {Ponman}, {Bower},
  {Booth}, {Dalla Vecchia}, {Crain}, {Springel}, {Theuns}, \&
  {Wiersma}}]{mccarthy10}
{McCarthy}, I.~G., {Schaye}, J., {Ponman}, T.~J., {et~al.} 2010, MNRAS, 406,
  822

\bibitem[{{McGee} \& {Balogh}(2010)}]{mcgee10}
{McGee}, S.~L., \& {Balogh}, M.~L. 2010, MNRAS, 403, L79

\bibitem[{{McLaughlin}(1999)}]{mcl99}
{McLaughlin}, D.~E. 1999, AJ, 117, 2398

\bibitem[{Muanwong {et~al.}(2001)Muanwong, Thomas, Kay, Pearce, \&
  Couchman}]{muanwong01}
Muanwong, O., Thomas, P.~A., Kay, S.~T., Pearce, F.~R., \& Couchman, H.~M.~P.
  2001, ApJ, 552, L27

\bibitem[{{Nagai} \& {Lau}(2011)}]{nagai11}
{Nagai}, D., \& {Lau}, E.~T. 2011, ApJ, 731, L10

\bibitem[{{Nagai} {et~al.}(2007){Nagai}, {Vikhlinin}, \& {Kravtsov}}]{nagai07}
{Nagai}, D., {Vikhlinin}, A., \& {Kravtsov}, A.~V. 2007, ApJ, 655, 98

\bibitem[{Navarro {et~al.}(1995)Navarro, Frenk, \& White}]{navarro95}
Navarro, J.~F., Frenk, C.~S., \& White, S.~D.~M. 1995, MNRAS, 275, 720

\bibitem[{{Osmond} \& {Ponman}(2004)}]{osmond04}
{Osmond}, J.~P.~F., \& {Ponman}, T.~J. 2004, MNRAS, 350, 1511

\bibitem[{{Piffaretti} \& {Valdarnini}(2008)}]{piffaretti08}
{Piffaretti}, R., \& {Valdarnini}, R. 2008, A\&A, 491, 71

\bibitem[{{Rasia} {et~al.}(2006){Rasia}, {Ettori}, {Moscardini}, {Mazzotta},
  {Borgani}, {Dolag}, {Tormen}, {Cheng}, \& {Diaferio}}]{rasia06}
{Rasia}, E., {Ettori}, S., {Moscardini}, L., {et~al.} 2006, MNRAS, 369, 2013

\bibitem[{{Rasmussen} \& {Ponman}(2004)}]{rasmussen04}
{Rasmussen}, J., \& {Ponman}, T.~J. 2004, MNRAS, 349, 722

\bibitem[{{Rasmussen} {et~al.}(2006){Rasmussen}, {Ponman}, {Mulchaey}, {Miles},
  \& {Raychaudhury}}]{rasmussen06b}
{Rasmussen}, J., {Ponman}, T.~J., {Mulchaey}, J.~S., {Miles}, T.~A., \&
  {Raychaudhury}, S. 2006, MNRAS, 373, 653

\bibitem[{{Rykoff} {et~al.}(2008{\natexlab{a}}){Rykoff}, {McKay}, {Becker},
  {Evrard}, {Johnston}, {Koester}, {Rozo}, {Sheldon}, \&
  {Wechsler}}]{rykoff08a}
{Rykoff}, E.~S., {McKay}, T.~A., {Becker}, M.~R., {et~al.} 2008{\natexlab{a}},
  ApJ, 675, 1106

\bibitem[{{Rykoff} {et~al.}(2008{\natexlab{b}}){Rykoff}, {Evrard}, {McKay},
  {Becker}, {Johnston}, {Koester}, {Nord}, {Rozo}, {Sheldon}, {Stanek}, \&
  {Wechsler}}]{rykoff08b}
{Rykoff}, E.~S., {Evrard}, A.~E., {McKay}, T.~A., {et~al.} 2008{\natexlab{b}},
  MNRAS, 387, L28

\bibitem[{{Sanderson} {et~al.}(2009{\natexlab{a}}){Sanderson}, {Edge}, \&
  {Smith}}]{san09b}
{Sanderson}, A.~J.~R., {Edge}, A.~C., \& {Smith}, G.~P. 2009{\natexlab{a}},
  MNRAS, 398, 1698

\bibitem[{{Sanderson} {et~al.}(2009{\natexlab{b}}){Sanderson}, {O'Sullivan}, \&
  {Ponman}}]{san09}
{Sanderson}, A.~J.~R., {O'Sullivan}, E., \& {Ponman}, T.~J. 2009{\natexlab{b}},
  MNRAS, 395, 764

\bibitem[{{Sanderson} \& {Ponman}(2010)}]{san10}
{Sanderson}, A.~J.~R., \& {Ponman}, T.~J. 2010, MNRAS, 402, 65

\bibitem[{{Sanderson} {et~al.}(2003){Sanderson}, {Ponman}, {Finoguenov},
  {Lloyd-Davies}, \& {Markevitch}}]{san03}
{Sanderson}, A.~J.~R., {Ponman}, T.~J., {Finoguenov}, A., {Lloyd-Davies},
  E.~J., \& {Markevitch}, M. 2003, MNRAS, 340, 989

\bibitem[{{Sanderson} {et~al.}(2006){Sanderson}, {Ponman}, \&
  {O'Sullivan}}]{san06}
{Sanderson}, A.~J.~R., {Ponman}, T.~J., \& {O'Sullivan}, E. 2006, MNRAS, 372,
  1496

\bibitem[{{Short} \& {Thomas}(2009)}]{short09}
{Short}, C.~J., \& {Thomas}, P.~A. 2009, ApJ, 704, 915

\bibitem[{{Sivanandam} {et~al.}(2009){Sivanandam}, {Zabludoff}, {Zaritsky},
  {Gonzalez}, \& {Kelson}}]{sivanandam09}
{Sivanandam}, S., {Zabludoff}, A.~I., {Zaritsky}, D., {Gonzalez}, A.~H., \&
  {Kelson}, D.~D. 2009, ApJ, 691, 1787

\bibitem[{{Skrutskie} {et~al.}(2006)}]{skrutskie06}
{Skrutskie}, M.~F., {et~al.} 2006, AJ, 131, 1163

\bibitem[{{Snowden} {et~al.}(2004){Snowden}, {Collier}, \& {Kuntz}}]{snowden04}
{Snowden}, S.~L., {Collier}, M.~R., \& {Kuntz}, K.~D. 2004, ApJ, 610, 1182

\bibitem[{{Sun} {et~al.}(2009){Sun}, {Voit}, {Donahue}, {Jones}, {Forman}, \&
  {Vikhlinin}}]{sun09}
{Sun}, M., {Voit}, G.~M., {Donahue}, M., {et~al.} 2009, ApJ, 693, 1142

\bibitem[{{Thode}(2002)}]{thode02}
{Thode}, H.~C. 2002, {Testing for Normality} (CRC Press)

\bibitem[{{Urban} {et~al.}(2011){Urban}, {Werner}, {Simionescu}, {Allen}, \&
  {B{\"o}hringer}}]{urban11}
{Urban}, O., {Werner}, N., {Simionescu}, A., {Allen}, S.~W., \&
  {B{\"o}hringer}, H. 2011, MNRAS, 414, 2101

\bibitem[{{van Daalen} {et~al.}(2011){van Daalen}, {Schaye}, {Booth}, \& {Dalla
  Vecchia}}]{vanDaalen11}
{van Daalen}, M.~P., {Schaye}, J., {Booth}, C.~M., \& {Dalla Vecchia}, C. 2011,
  MNRAS, 415, 3649

\bibitem[{Venables \& Ripley(2002)}]{venables02}
Venables, W.~N., \& Ripley, B.~D. 2002, Modern Applied Statistics with S, 4th
  edn. (New York: Springer), iSBN 0-387-95457-0

\bibitem[{{Vikhlinin} {et~al.}(2007){Vikhlinin}, {Burenin}, {Forman}, {Jones},
  {Hornstrup}, {Murray}, \& {Quintana}}]{vikhlinin07}
{Vikhlinin}, A., {Burenin}, R., {Forman}, W.~R., {et~al.} 2007, in Heating
  versus Cooling in Galaxies and Clusters of Galaxies, ed. H.~{B{\"o}hringer},
  G.~W. {Pratt}, A.~{Finoguenov}, \& P.~{Schuecker}, 48

\bibitem[{{Vikhlinin} {et~al.}(1999){Vikhlinin}, {Forman}, \&
  {Jones}}]{vikhlinin99}
{Vikhlinin}, A., {Forman}, W., \& {Jones}, C. 1999, ApJ, 525, 47

\bibitem[{{Vikhlinin} {et~al.}(2006){Vikhlinin}, {Kravtsov}, {Forman}, {Jones},
  {Markevitch}, {Murray}, \& {Van Speybroeck}}]{vikhlinin06}
{Vikhlinin}, A., {Kravtsov}, A., {Forman}, W., {et~al.} 2006, ApJ, 640, 691

\bibitem[{{Voit} \& {Bryan}(2001)}]{voit01}
{Voit}, G.~M., \& {Bryan}, G.~L. 2001, Nature, 414, 425

\bibitem[{Wickham(2009)}]{wickham09}
Wickham, H. 2009, ggplot2: elegant graphics for data analysis (Springer New
  York), iSBN 978-0-387-98140-6

\bibitem[{{Zaritsky} {et~al.}(2006){Zaritsky}, {Gonzalez}, \&
  {Zabludoff}}]{zaritsky06}
{Zaritsky}, D., {Gonzalez}, A.~H., \& {Zabludoff}, A.~I. 2006, ApJ, 638, 725

\bibitem[{{Zibetti} {et~al.}(2005){Zibetti}, {White}, {Schneider}, \&
  {Brinkmann}}]{zibetti05}
{Zibetti}, S., {White}, S.~D.~M., {Schneider}, D.~P., \& {Brinkmann}, J. 2005,
  MNRAS, 358, 949

\end{thebibliography}
\label{lastpage}

\end{document}